\let\accentvec\vec
\documentclass[english,longauth]{aa}
\usepackage{savesym}
\usepackage{longtable}
\usepackage[utf8]{inputenc}
\usepackage[T1]{fontenc}
\usepackage[]{natbib}
\usepackage{lmodern} 
\usepackage{graphicx}
\usepackage{float}
\usepackage{tabularx}
\usepackage{color}
\usepackage{ulem}
\usepackage{booktabs}
\usepackage{mathabx}
\savesymbol{second}
\savesymbol{degree}
\usepackage{multirow}
\usepackage{placeins}
\usepackage{url}
\usepackage[squaren, Gray, cdot]{SIunits}
\restoresymbol{SInits}{second}
\restoresymbol{SInits}{degree}
\usepackage[colorlinks=true,urlcolor=blue,linkcolor=blue,citecolor=blue,breaklinks=true]{hyperref}

\makeatletter

\let\vec\accentvec

\usepackage{natbib,twoopt}
\bibpunct{(}{)}{;}{a}{}{,}             
\makeatletter
  \newcommandtwoopt{\citeads}[3][][]{\href{http://adsabs.harvard.edu/abs/#3}%
    {\def\hyper@linkstart##1##2{}%
     \let\hyper@linkend\@empty\citealp[#1][#2]{#3}}}
  \newcommandtwoopt{\citepads}[3][][]{\href{http://adsabs.harvard.edu/abs/#3}%
    {\def\hyper@linkstart##1##2{}%
     \let\hyper@linkend\@empty\citep[#1][#2]{#3}}}
  \newcommandtwoopt{\citetads}[3][][]{\href{http://adsabs.harvard.edu/abs/#3}%
    {\def\hyper@linkstart##1##2{}%
     \let\hyper@linkend\@empty\citet[#1][#2]{#3}}}
  \newcommandtwoopt{\citeyearads}[3][][]%
    {\href{http://adsabs.harvard.edu/abs/#3}
    {\def\hyper@linkstart##1##2{}%
     \let\hyper@linkend\@empty\citeyear[#1][#2]{#3}}}
\makeatother

\makeatletter

\providecommand{\LyX}{L\kern-.1667em\lower.25em\hbox{Y}\kern-.125emX\@}

\DeclareRobustCommand*{\lyxarrow}{%
\@ifstar
{\leavevmode\,$\triangleleft$\,\allowbreak}
{\leavevmode\,$\triangleright$\,\allowbreak}}

\makeatother

\begin{document}

\title{A transit survey to search for planets around hot subdwarfs: I. methods and performance tests on light curves from Kepler, K2, TESS, and CHEOPS}

\author{
V. Van Grootel\inst{1}
\and F.~J. Pozuelos\inst{1,2}
\and A. Thuillier\inst{1}
\and S. Charpinet\inst{3}
\and L. Delrez\inst{1,2,4}
\and M. Beck\inst{4}
\and A. Fortier\inst{5,6}
\and S. Hoyer\inst{7}
\and S.~G. Sousa\inst{8}
\and B.~N. Barlow\inst{9}
\and N. Billot\inst{4}
\and M. D\'evora-Pajares\inst{10,11}
\and R.~H. \O stensen\inst{12}
\and Y. Alibert\inst{5}
\and R. Alonso\inst{13,14}
\and G. Anglada Escud\'e\inst{15,16}
\and J. Asquier\inst{17}
\and D. Barrado\inst{18}
\and S.~C.~C. Barros\inst{8,19}
\and W. Baumjohann\inst{20}
\and T. Beck\inst{5}
\and A. Bekkelien\inst{4}
\and W. Benz\inst{5,6}
\and X. Bonfils\inst{21}
\and A. Brandeker\inst{22}
\and C. Broeg\inst{5}
\and G. Bruno\inst{23}
\and T. B\'arczy\inst{24}
\and J. Cabrera\inst{25}
\and A.~C. Cameron\inst{26}
\and S. Charnoz\inst{27}
\and M.~B. Davies\inst{28}
\and M. Deleuil\inst{7}
\and O.~D.~S. Demangeon\inst{8,19}
\and B.-O. Demory\inst{5}
\and D. Ehrenreich\inst{4}
\and A. Erikson\inst{25}
\and L. Fossati\inst{20}
\and M. Fridlund\inst{29,30}
\and D. Futyan\inst{4}
\and D. Gandolfi\inst{31}
\and M. Gillon\inst{2}
\and M. Guedel\inst{32}
\and K. Heng\inst{6,33}
\and K.~G. Isaak\inst{17}
\and L. Kiss\inst{34,35,36}
\and J. Laskar\inst{37}
\and A. Lecavelier des Etangs\inst{38}
\and M. Lendl\inst{4}
\and C. Lovis\inst{4}
\and D. Magrin\inst{39}
\and P.~F.~L. Maxted\inst{40}
\and M. Mecina\inst{32,20}
\and A.~J. Mustill\inst{28}
\and V. Nascimbeni\inst{39}
\and G. Olofsson\inst{22}
\and R. Ottensamer\inst{32}
\and I. Pagano\inst{23}
\and E. Pall\'e\inst{13,14}
\and G. Peter\inst{41}
\and G. Piotto\inst{42,39}
\and J.-Y. Plesseria\inst{43}
\and D. Pollacco\inst{33}
\and D. Queloz\inst{4,44}
\and R. Ragazzoni\inst{42,39}
\and N. Rando\inst{17}
\and H. Rauer\inst{25,45,46}
\and I. Ribas\inst{15,16}
\and N.~C. Santos\inst{8,19}
\and G. Scandariato\inst{23}
\and D. S\'egransan\inst{4}
\and R. Silvotti\inst{47}
\and A.~E. Simon\inst{5}
\and A.~M.~S. Smith\inst{25}
\and M. Steller\inst{20}
\and G. M. Szab{\'o}\inst{48,49}
\and N. Thomas\inst{5}
\and S. Udry\inst{4}
\and V. Viotto\inst{39}
\and N.~A. Walton\inst{50}
\and K. Westerdorff\inst{41}
\and T.~G. Wilson\inst{26}
}

\institute{Space sciences, Technologies and Astrophysics Research (STAR) Institute, Universit\'e
  de Li\`ege, 19C All\'ee du 6 Ao\^ut, B-4000 Li\`ege, Belgium \\
\email{valerie.vangrootel@uliege.be}
\and{Astrobiology Research Unit, Universit\'e de Li\`ege, All\'ee du 6 Ao\^ut 19C, B-4000 Li\`ege, Belgium}
\and{Institut de Recherche en Astrophysique et Plan\'etologie, CNRS, Universit\'e de Toulouse, CNES, 14 avenue Edouard Belin, F-31400 Toulouse, France}
\and{Observatoire Astronomique de l'Universit\'e de Gen\`eve, Chemin Pegasi 51, Versoix, Switzerland}
\and{Physikalisches Institut, University of Bern, Gesellsschaftstrasse 6, 3012 Bern, Switzerland}
\and{Center for Space and Habitability, Gesellsschaftstrasse 6, 3012 Bern, Switzerland}
\and{Aix Marseille Univ, CNRS, CNES, LAM, Marseille, France}
\and{Instituto de Astrof\'isica e Ci\^encias do Espa\c{c}o, Universidade do Porto, CAUP, Rua das Estrelas, 4150-762 Porto, Portugal}
\and{Department of Physics, High Point University, One University Parkway, High Point, NC 27268}
\and{Universidad Internacional de Valencia (VIU), Carrer del Pintor Sorolla 21, 46002, Valencia, Spain}
\and{Dpto. Física Teórica y del Cosmos, Universidad de Granada, 18071, Granada, Spain}
\and{Department of Physics, Astronomy and Materials Science, Missouri State University, 901 S. National, Springfield,
MO 65897, USA}
\and{Instituto de Astrof\'\i sica de Canarias, 38200 La Laguna, Tenerife, Spain}
\and{Departamento de Astrof\'\i sica, Universidad de La Laguna, 38206 La Laguna, Tenerife, Spain}
\and{Institut de Ci\`encies de l'Espai (ICE, CSIC), Campus UAB, Can Magrans s/n, 08193 Bellaterra, Spain}
\and{Institut d'Estudis Espacials de Catalunya (IEEC), 08034 Barcelona, Spain}
\and{ESTEC, European Space Agency, 2201AZ, Noordwijk, NL}
\and{Depto. de Astrofísica, Centro de Astrobiologia (CSIC-INTA), ESAC campus, 28692 Villanueva de la Cãda (Madrid), Spain}
\and{Departamento de F\'isica e Astronomia, Faculdade de Ci\^encias, Universidade do Porto, Rua do Campo Alegre, 4169-007 Porto, Portugal}
\and{Space Research Institute, Austrian Academy of Sciences, Schmiedlstrasse 6, A-8042 Graz, Austria}
\and{Université Grenoble Alpes, CNRS, IPAG, 38000 Grenoble, France}
\and{Department of Astronomy, Stockholm University, AlbaNova University Center, 10691 Stockholm, Sweden}
\and{INAF, Osservatorio Astrofisico di Catania, Via S. Sofia 78, 95123 Catania, Italy}
\and{Admatis, Miskok, Hungary}
\and{Institute of Planetary Research, German Aerospace Center (DLR), Rutherfordstrasse 2, 12489 Berlin, Germany}
\and{Centre for Exoplanet Science, SUPA School of Physics and Astronomy, University of St Andrews, North Haugh, St Andrews KY16 9SS, UK}
\and{Universit\'e de Paris, Institut de physique du globe de Paris, CNRS, F-75005 Paris, France}
\and{Lund Observatory, Dept. of Astronomy and Theoretical Physics, Lund University, Box 43, 22100 Lund, Sweden}
\and{Leiden Observatory, University of Leiden, PO Box 9513, 2300 RA Leiden, The Netherlands}
\and{Department of Space, Earth and Environment, Chalmers University of Technology, Onsala Space Observatory, 43992 Onsala, Sweden}
\and{Dipartimento di Fisica, Universit\`a degli Studi di Torino, via Pietro Giuria 1, I-10125, Torino, Italy}
\and{University of Vienna, Department of Astrophysics, Türkenschanzstrasse 17, 1180 Vienna, Austria}
\and{Department of Physics, University of Warwick, Gibbet Hill Road, Coventry CV4 7AL, United Kingdom}
\and{Konkoly Observatory, Research Centre for Astronomy and Earth Sciences, 1121 Budapest, Konkoly Thege Miklós út 15-17, Hungary}
\and{ELTE E\"otv\"os Lor\'and University, Institute of Physics, P\'azm\'any P\'eter s\'et\'any 1/A, 1117 Budapest, Hungary}
\and{Sydney Institute for Astronomy, School of Physics A29, University of Sydney, NSW 2006, Australia}
\and{IMCCE, UMR8028 CNRS, Observatoire de Paris, PSL Univ., Sorbonne Univ., 77 av. Denfert-Rochereau, 75014 Paris, France}
\newpage
\and Institut d'astrophysique de Paris, UMR7095 CNRS, Universit\'e Pierre \& Marie Curie, 98bis blvd. Arago, 75014 Paris, France
\and{INAF, Osservatorio Astronomico di Padova, Vicolo dell'Osservatorio 5, 35122 Padova, Italy}
\and{Astrophysics Group, Keele University, Staffordshire, ST5 5BG, United Kingdom}
\and{Institute of Optical Sensor Systems, German Aerospace Center (DLR), Rutherfordstrasse 2, 12489 Berlin, Germany}
\and{Dipartimento
di Fisica e Astronomia, Universit\`a degli Studi di Padova, 
Vicolo dell'Osservatorio 3, 35122 Padova, Italy}
\and{Centre Spatial de Li\`ege, STAR institute, Universit\'e de Li\`ege, avenue du Pr\'e Aily, B-4031 Angleur (Li\`ege), Belgium}
\and{Cavendish Laboratory, JJ Thomson Avenue, Cambridge CB3 0HE, UK}
\and{Center for Astronomy and Astrophysics, Technical University Berlin, Hardenberstrasse 36, 10623 Berlin, Germany}
\and{Institut für Geologische Wissenschaften, Freie Universität Berlin, 12249 Berlin, Germany}
\and{INAF-Osservatorio Astrofisico di Torino, Strada dell’Osservatorio 20, 10025 Pino Torinese, Italy}
\and{ELTE Eötvös Loránd University, Gothard Astrophysical Observatory, 9700 Szombathely, Szent Imre h. u. 112, Hungary}
\and{MTA-ELTE Exoplanet Research Group, 9700 Szombathely, Szent Imre h. u. 112, Hungary}
\and{Institute of Astronomy, University of Cambridge, Madingley Road, Cambridge, CB3 0HA, United Kingdom}
}

\offprints{V. Van Grootel}

\date{Received ...; Accepted...}

\abstract
{Hot subdwarfs experienced strong mass loss on the Red Giant Branch (RGB) and are now hot and small He-burning objects. Such stars constitute excellent opportunities to address the question of the evolution of exoplanetary systems directly after the RGB phase of evolution.}
{In this project we aim to perform a transit survey in all available light curves of hot subdwarfs from space-based telescopes (Kepler, K2, TESS, and CHEOPS), with our custom-made pipeline {\fontfamily{pcr}\selectfont SHERLOCK}, in order to determine the occurrence rate of planets around these stars, as a function of orbital period and planetary radius. We also aim to determine whether planets previously engulfed in the envelope of their red giant host star can survive, even partially, as a planetary remnant.} 
{In this first paper, we perform injection-and-recovery tests of synthetic transits for a selection of representative Kepler, K2 and TESS light curves, to determine which transiting bodies, in terms of object radius and orbital period, we will be able to detect with our tools. We also provide such estimates for CHEOPS data, which we analyze with the {\tt{pycheops}} package.
}
{Transiting objects with a radius $\lesssim$ 1.0 $R_{\Earth}$ can be detected in most of Kepler, K2 and CHEOPS targets for the shortest orbital periods (1~d and below), reaching values as small as $\sim$0.3 $R_{\Earth}$ in the best cases. Reaching sub-Earth-sized bodies is achieved only for the brightest  TESS targets, and the ones observed during a significant number of sectors. We also give a series of representative results for farther and bigger planets, for which the performances strongly depend on the target magnitude, the length and the quality of the data.}
{The TESS sample will provide the most important statistics for the global aim of measuring the planet occurrence rate around hot subdwarfs. The Kepler, K2 and CHEOPS data will allow us to search for planetary remnants, i.e. very close and small (possibly disintegrating) objects, which would have partly survived the engulfment in their red giant host.}


\titlerunning{A transit survey to search for planets around hot subdwarfs: I.}
\authorrunning{V. Van Grootel et al.}

\maketitle

\section{Introduction}
Hot subdwarf B (sdB) stars are hot and compact stars ($T_{\rm eff} = $ 20 000 $-$ 40 000 K and log $g = $ 5.2 to 6.2; \citealt{1994ApJ...432..351S}), lying on the blue tail of the horizontal branch (HB), i.e., the extreme horizontal branch (EHB). The HB stage corresponds to core-He burning objects and follows the red giant branch (RGB) phase. Unlike most post-RGB stars that cluster at the red end of the HB (the so-called red clump, RC) because they lose almost no envelope on the RGB \citep{2012MNRAS.419.2077M}, sdB stars experienced strong mass loss on the RGB and have extremely thin residual H envelopes ($M_{\rm env} < 0.01 M_{\odot}$, \citealt{1986A&A...155...33H}). This extremely thin envelope explains their high effective temperatures and their inability to sustain H-shell burning. This prevents these stars from ascending the asymptotic giant branch (AGB) after core-He exhaustion \citep{1993ApJ...419..596D}. About 60\% of sdBs reside in binary systems, with about half of them being in close binaries with orbital periods up to a few days \citep[see, e.g.,][]{1994AJ....107.1565A,2001MNRAS.326.1391M}, while the other half reside in wider binaries with orbital periods up to several years \citep{2003AJ....126.1455S,2018MNRAS.473..693V}. Binary interactions (through common-envelope -CE- evolution for the short orbits, and stable Roche lobe overflow -RLOF- evolution for the wide orbits) are therefore the main culprit for this extreme mass loss \citep{2002MNRAS.336..449H,2003MNRAS.341..669H}. The hot O-type subdwarfs, or sdO stars, have $T_{\rm eff} = $ 40 000 $-$ 80 000 K and a wide range of surface gravities. The compact sdO stars (log $g = 5.2-6.2$) are either post-EHB objects, or direct post-RGB objects (through a so-called late hot He-flash; \citealt{2008A&A...491..253M}), or end products of merger events \citep{1990ApJ...353..215I,2000MNRAS.313..671S,2002MNRAS.333..121S}. The sdOs with log g < 5.2 are post-AGB stars, i.e. stars that have ascended the giant branch a second time after core-He burning exhaustion \citep{2016A&A...587A.101R}. 

The formation of the $\sim$40\% of sdB stars that appear to be single has been a mystery for decades. In the absence of a companion, it is hard to explain how the star can expel most of its envelope on the RGB and still achieve core-He burning ignition. Recently, \citet{2020A&A...642A.180P} suggested to all sdB stars might originate from binary evolution. Merger scenarios involving two low-mass white dwarfs have also been investigated \citep{1984ApJ...277..355W,2002MNRAS.336..449H,2003MNRAS.341..669H,2012MNRAS.419..452Z}, but several facts challenge this hypothesis. Firstly, such compact low-mass white dwarf binaries are quite rare, even if some candidates are identified \citep{2019ApJ...883...51R}. Secondly, the mass distribution of single and binary sdB stars are indistinguishable \citep[][Table 3 in particular]{2012A&A...539A..12F}. This mass distribution is mainly obtained from asteroseismology (some sdB stars exhibit oscillations, which allow the precise and accurate determination of the stellar parameters, including total mass; \citealt{2013A&A...553A..97V}), as well as from binary light curve modeling for hot subdwarfs in eclipsing binary systems. Single and binary mass distributions peak at $\sim0.47 M_{\odot}$, which is the minimum mass required to ignite He through a He-flash at the tip of RGB (stars of $\gtrsim 2.3 M_{\odot}$ are able to ignite He quietly at lower core masses, down to $\sim 0.33 M_{\odot}$, but the more massive the stars, the rarer they are). A mass distribution of single sdB stars from mergers would on the contrary be much broader (0.4$-$0.7 $M_{\odot}$; \citealt{2002MNRAS.336..449H}). With the DR2 release of \textit{Gaia} \citep{2018A&A...616A...1G} and precise distances for many hot subdwarfs \citep{2020A&A...635A.193G}, it is also now possible to build a spectrophotometric mass distribution for a much bigger sample compared to what was achieved with the hot subdwarf pulsators or those in eclipsing binaries only. Individual masses are much less precise compared those obtained by asteroseismology or binary light curve modeling \citep{schneider2019zenodo}. However, also here single and binary spectrophotometric mass distributions share the same properties, which tend to discard the hypothesis of different origins for single and binary sdB stars. The third piece of evidence against merger scenarios (that would most likely give fast-rotating objects) is the very slow rotation of almost all single sdB stars, as obtained through $v\sin i$ measurements \citep{2012A&A...543A.149G} or from asteroseismology \citep{2018OAst...27..112C}. Moreover, their rotation rates are in direct line with the core rotation rates observed in RC stars \citep{2012A&A...548A..10M}, which is another strong indication that these stars and the single sdB stars do share a same origin, i.e., they are post-RGB stars.

The question of the evolution of exoplanet systems after the main sequence of their host is generally addressed by studying exoplanets around subgiants, RGB stars and normal HB (RC) stars (hereafter the 'classical' evolved stars). These 'classical' evolved stars are typically very large stars, with radii ranging from $\sim 5-10~R_{\odot}$ to more than 1000~$R_{\odot}$, much larger than hot subdwarfs with radii in the range $\sim 0.1-0.3~R_{\odot}$ \citep{2016PASP..128h2001H}, and their mass is typically larger than $\sim$ 1.5~$M_{\odot}$, compared to $\sim0.47 M_{\odot}$ for hot subdwarfs. Both the transit and radial velocity (RV) methods are challenging for these 'classical' evolved stars due to transit depth dilution and additional noise sources \citep{2016AJ....152..143V}. Another difficulty for the question of the fate of exoplanet systems after the RGB phase itself is the difficulty on distinguishing RGB and RC stars based on their spectroscopic parameters alone, and sometimes even with help of asteroseismology \citep{2019ApJ...885...31C}. As a consequence, only large/massive planets are detected around the 'classical' evolved stars \citep[][and references therein]{2020arXiv200601277J}. A dearth of close-in giant planets is observed around these evolved stars compared to solar-type main sequence stars \citep{2008PASJ...60.1317S,2009A&A...505.1311D}. This may be caused by planet engulfment by the host star, but with current technologies it is not possible to determine if smaller planets and remnants (such as the dense core of former giant planets) are present. The lack of close-in giant planets may also be explained by intrinsically different planetary formation for these intermediate-mass stars \citep[see the discussion in][]{2020arXiv200601277J}. Ultimately, the very existence of planet remnants may be linked to the ejection of most of the envelope on the RGB that occur for hot subdwarfs, while for 'classical' evolved stars, nothing stops the in-spiraling planet inside the host star, and in all cases the planet finally merges with the star, or is fully tidally disrupted, or totally ablated by heating or by the strong stellar wind. In other words, the ejection of the envelope not only makes possible the detection of small objects as remnants, but most importantly may actually be the cause of the existence of such remnants, by stopping the spiraling in the host star.

Many studies focused on white dwarfs (the ultimate fate of $\sim$97\% of all stars), including the direct observations of transiting disintegrating planetesimals \citep{2015Natur.526..546V}, the accretion of a giant planet \citep{2019Natur.576...61G}, and, most recently, the transit of a giant planet \citep{2020Natur.585..363V}. More than 25\% of all single white dwarfs exhibit metal pollution in their atmospheres (which should be pure H or He due to gravitational settling of heavier elements in these very high surface gravity objects), which is generally interpreted as accretion of surrounding planetary remnants material \citep[][and references therein]{2018MNRAS.477...93H}. Statistics on the occurrence rate of planets around white dwarfs as a function of orbital period and planet radius have also been established \citep{2014ApJ...796..114F,2018MNRAS.474.4603V,2019MNRAS.487..133W}. However, the vast majority of white dwarfs experienced two giant phases of evolution, namely the RGB and the AGB. The AGB expansion and strong mass loss, followed by the planetary-nebula phase, will have a profound effect on the orbital stability of surrounding bodies \citep[e.g.][]{2002ApJ...572..556D,2014MNRAS.437.1404M,2021MNRAS.501L..43M}. Hence no direct conclusion concerning the effect of RGB expansion alone on the exoplanet systems can be drawn from white dwarfs.

Hot subdwarfs therefore constitute excellent opportunities to address the question of the evolution of exoplanet systems after the RGB phase of evolution. It is precisely this potential we aim to exploit in this project, by determining the occurrence rate of planets around hot subdwarfs, as a function of orbital period and planet radius. We will achieve this objective by performing a transit search in all available light curves of hot subdwarfs from space-based observatories - such as Kepler \citep{2010Sci...327..977B}, K2 \citep{2014PASP..126..398H}, TESS \citep{2014SPIE.9143E..20R}, and CHEOPS \citep{2020ExA...tmp...53B}. In this first paper, we provide in Sect. \ref{review} a review of the current status of the search of planets around hot subdwarfs with the different detection methods. We present the observations and the tools used to perform our transit search in Sect. \ref{tools}. We provide extensive tests of the photometric quality of the light curves in Sect. \ref{inj} and \ref{Perf_CHEOPS}, and we give our conclusions and future work in Sect. \ref{cc}. 

\section{Search for planets around hot subdwarfs: current status}
\label{review}
To date, several planet detections around hot subdwarfs have been claimed, but none of them received confirmation. Using the pulsation timing method (variation of the oscillation periods of sdB pulsators), planets of a few Jupiter masses in orbits about 1 AU were announced around V391 Peg and DW Lyn \citep{2007Natur.449..189S,2012AN....333.1099L}, but these claims have recently been refuted \citep{2018A&A...611A..85S,2020A&A...638A.108M}. Based on weak signals interpreted as reflection and thermal re-emission in Kepler light curves, five very close-in (with orbital periods of a few hours) Earth-sized planets have been claimed to orbit KIC 05807616 \citep{2011Natur.480..496C} and KIC 10001893 \citep{2014A&A...570A.130S}. However, the attribution of such signals to exoplanets is debatable \citep{2015A&A...581A...7K,2019A&A...627A..86B}. Using the RV method, \citet{2009ApJ...702L..96G} announced the discovery of a close-in ($P_{\rm orb} =$ 2.4 days) planet of several Jupiter masses around HD 149382, but it was ruled out by high-precision RV measurements obtained with the Hobby-Eberly Telescope spectrograph, excluding the presence of almost any substellar companion with $P_{\rm orb} <$ 28 days and $M \sin i \gtrsim 1 M_{\rm Jup}$ \citep{2011ApJ...743...88N}. No close massive planets (down to a few Jupiter masses) were found from a mini RV survey carried out with the HARPS-N spectrograph on 8 apparently single hot subdwarfs \citep{2020arXiv200204545S}. 

Several ground-based surveys, with both photometric and RV techniques, target the red dwarf or brown dwarf close companions to hot subdwarfs \citep{2018A&A...614A..77S,2019A&A...630A..80S}. Such companions are frequent \citep{2018A&A...614A..77S}, but no Jupiter-like planets have been found to date. In contrast, several discoveries of circumbinary massive planets have been announced in close, post-CE evolution sdB+dM eclipsing systems through eclipse timing variations, e.g. HW Vir, the prototype of the class \citep{2009AJ....137.3181L,2012A&A...543A.138B}, NSVS 14256825 \citep{2019RAA....19..134Z}, HS 0705+6700 \citep{2015JBAA..125..284P}, NY Vir \citep{2012ApJ...745L..23Q}, and 2M 1938+4603 \citep{2015A&A...577A.146B}. Such planets could correspond to first generation, second generation \citep{2014A&A...563A..61S,2014A&A...562A..19V}, or hybrid planets (which are formed from ejected stellar material accreted on remnants of first generation planets; \citealt{2013A&A...549A..95Z}). Among the ten well studied HW Vir systems, all but one show eclipse timing variations \citep{2016PASP..128h2001H,2018haex.bookE..96M}. This may call for another explanation than planets (perhaps something analog to the mechanism suggested for eclipse timing variations in white dwarf binaries; \citealt{2016MNRAS.460.3873B}), since the occurrence of circumbinary planets around close main-sequence binaries that are the progenitors of such systems is $\sim$1\% only \citep{2012Natur.481..475W}. The properties of the claimed planets often change or are discarded after new measurements \citep{2016PASP..128h2001H,2018haex.bookE..96M}, while the orbits are regularly found to be dynamically unstable \citep[e.g.][]{2013MNRAS.431.2150W}. None of these claimed circumbinary planets has been confirmed through another technique.

\section{Observations and methods}
\label{tools}
\subsection{Space-based light curves of hot subdwarfs}

In the original Kepler field, 72 hot subdwarfs were observed at the short cadence (SC) of 1 minute for at least one quarter, including the commissioning quarter Q0 that started on May 2, 2009. During the one-year survey phase that followed Q0 (quarters Q1 of 33.5 days, and Q2 to Q4 of 90 days each, divided in monthly subquarters), 15 sdB stars were found to pulsate \citep{2010MNRAS.409.1470O,2011MNRAS.414.2860O}. These 15 stars were consequently observed for the rest of the mission at SC (with exceptions of some quarters for a couple of sdB pulsators, see details in Table \ref{sd_Kepler}). Three other sdB pulsators, known as B3, B4 and B5, were found in the open cluster NGC 6791 \citep{2011ApJ...740L..47P} and were observed in SC for various durations (see Table \ref{sd_Kepler}). Among non-pulsators, 47 hot subdwarfs of the B type (sdB and sdOB) were observed for at least one month at SC (5 of them for several quarters), as well as 7 sdO stars. At the long cadence (LC) of 30 minutes, these 54 non-pulsating hot subdwarfs were generally observed for several quarters, and some of them for the whole duration of the mission. The list of hot subdwarf targets in the original Kepler field and details on the observing quarters in SC and LC can be found in Table \ref{sd_Kepler}. The primary Kepler mission stopped on May 11, 2013 during Q17.2, after the failure of a second reaction wheel necessary to stabilize the spacecraft and obtain the fine and stable pointing for observations of the original field.

The Kepler mission was then redesigned as K2, for which the two remaining reaction wheels allowed a stable pointing for $\sim$80 days of fields close to the ecliptic. An engineering test of 11 days in February 2014 confirmed the feasibility of this strategy, and 19 campaigns (campaign 0 to 18) were executed from March 2014 to July 2018, when exhaustion of propellants definitively ended the mission. In K2 fields, accounting only for confirmed hot subdwarfs, 39 sdB/sdOB pulsators were observed in SC through at least one campaign. Two more sdB pulsators were discovered through LC data only. 79 more sdB/sdOB non-pulsators, and 10 sdO non-pulsators, were also observed in SC. Finally, 44 hot subdwarfs were observed in LC only. Contrary to Kepler, K2 SC and LC data generally cover one campaign (of about 80 days duration) only, although a few number of stars were observed in two or three campaigns. The full list of hot subdwarfs observed by K2 and details can be found in Table \ref{sd_K2}.   

\begin{table}[!ht]
\caption{\label{sd_TESS_primary}Statistics on hot subdwarfs observed in the primary TESS mission (July 2018-July 2020).}
\begin{center}
\begin{tabular}{lc}
\hline\hline
Number of sectors & Number of stars   \tabularnewline
\hline
1 & 877 \tabularnewline
2 & 205 \tabularnewline
3 & 72  \tabularnewline
4& 23  \tabularnewline
5 & 21  \tabularnewline
6 & 24   \tabularnewline
7 & 7  \tabularnewline
8 & 10  \tabularnewline
9 & 6    \tabularnewline
10& 6   \tabularnewline
11& 13  \tabularnewline
12& 23 \tabularnewline
13 & 15  \tabularnewline
\hline \hline
G magnitude & Number of stars \tabularnewline
\hline
8-9 & 3 \tabularnewline
9-10 & 4 \tabularnewline
10-11& 18  \tabularnewline
11-12&60  \tabularnewline
12-13&162  \tabularnewline
13-14&278  \tabularnewline
14-15&384  \tabularnewline
15-16&341  \tabularnewline
16-17& 51  \tabularnewline
Beyond 17 & 1  \tabularnewline
\hline\hline
\end{tabular}
\end{center}
\end{table}

TESS (Transiting Exoplanet Survey Satellite) has been operational since July 2018. It is performing a high-precision photometric survey over almost the whole sky (about 90\%), avoiding only a narrow band around the ecliptic\footnote{\url{https://tess.mit.edu/observations/}}. The TESS primary two-year mission, which ended early July 2020, consisted of 26 sectors observed nearly continuously for $\sim$27.4 days each. Some overlap between sectors exists for the highest northern and southern ecliptic latitudes, hence some stars have been observed for several sectors (see Table \ref{sd_TESS_primary}). The primary mission TESS data products consist of SC observations sampled every 2 minutes for selected stars, as well as full-frame images (FFI) taken every 30 min containing data for all stars in the  field of view. Accounting for confirmed hot subdwarfs only, 1302 stars were observed for at least one sector at SC during primary mission. This list was assembled by the Working Group (WG) 8 on compact pulsators of the TESS Asteroseismic Consortium (TASC; see also \citealt{2019AJ....158..138S}). Table \ref{sd_TESS_primary} presents statistics about these TESS primary mission observations of hot subdwarfs, while the full list can be found at {\url{https://github.com/franpoz/Hot-Subdwarfs-Catalogues}} (see Appendix \ref{CDS_primary} for details). The TESS extended mission started on July 4, 2020, and revisits all sectors for the same duration, referred to with increasing numbers (Sector 27, 28, etc.). An 'ultra short cadence' of 20 s is now available in addition to the normal SC of 2 min, and FFIs are now taken every 10 minutes. After release of Sectors 27 to 31, 243 confirmed hot subdwarfs have been observed at 20 s cadence, and 670 more at 2 min cadence (these targets were also selected by WG8 of the TASC). Most of these targets were already observed in the primary mission (sectors 1 to 26), but there are also about one third of new targets that were not observed during the primary mission. The list can be found at {\url{https://github.com/franpoz/Hot-Subdwarfs-Catalogues}} (see Appendix \ref{CDS_extended} for details). It is expected that about 2300 hot subdwarfs will have been observed at the end of the two-year extended mission. 

\begin{table*}
\scriptsize
\caption{\label{CHEOPS_targets}List of hot subdwarf targets observed by CHEOPS (as of December 19, 2020).}
\begin{center}
\begin{tabular}{lccccc}
\hline\hline
\multirow{2}{*}{Name} & \multirow{2}{*}{Type} & \multirow{2}{*}{Gmag} &\# of orbits & Phase coverage & Min. planet size  \tabularnewline
&&&(as of 19 Nov 2020) & (days, for $>$80\% coverage)&($R_{\Earth}$, for SNR$=$5 and 0.18$R_{\odot}$ host)\tabularnewline
\hline
{\textbf{Active}}&&&&&\tabularnewline
HD 149382&sdB&8.80&14 (7x2)&0.47&0.4\tabularnewline
HD 127493&sdO&9.96&6.8 (2x1$+$4.8)&0.18&0.4 \tabularnewline
TYC981-1097-1&sd&12.01&18 (6x3)&0.68&0.7 \tabularnewline
Feige 110&sdOB&11.79&6 (3x2)&0.25&0.7 \tabularnewline
CW83-1419-09 &sdOB&12.04 & 12 (4x3) & 0.39 & 0.7 \tabularnewline
EC 14248-2647&sdOB&11.98&2 (1x2)&$<$0.10&0.7 \tabularnewline
PG 2219+094&sdB&11.90&5 (5x1)&0.18&0.7  \tabularnewline
PG 1352-023&sdOB&12.06&6 (3x2)&0.18&0.8 \tabularnewline
LS IV -12 1&sdO&11.11&4 (4x1)&0.18&0.8 \tabularnewline
Feige 14&sdB&12.77&5 (5x1)&0.11&0.8 \tabularnewline
EC 22081-1916&sdB&12.94&6 (3x2)&0.25&0.8 \tabularnewline
LS IV+06 2&He-sdO&12.14&5 (5x1)&0.18&0.8 \tabularnewline
MCT 2350-3026&sdO&12.07&8 (4x2)&0.32&0.8  \tabularnewline
TYC 982-614-1&sd&12.21&18 (6x3)&0.68&0.8 \tabularnewline
EC 20305-1417&sdB&12.34&6 (2x3)& 0.25 & 0.8 \tabularnewline
LS IV+109&He-sdO&11.97&10 (5x2)&0.39&0.8 \tabularnewline
PG 1432+004&sdB&12.75&4 (2x2)&0.11&0.8  \tabularnewline
TonS403&sdO&12.92&11 (11x1)& 0.25 &0.8 \tabularnewline
TYC 497-63-1&sdB&12.89&5 (5x1)& 0.11 &0.8\tabularnewline 
TYC 999-2458-1&sdB&12.59&3 (1x3)&0.18&0.9 \tabularnewline
TYC 499-2297-1&sdB&12.63&12 (6x2)&0.54&0.9 \tabularnewline
LS IV+00 21&sdOB&12.41&4 (2x2)& 0.18 &0.9 \tabularnewline
PG 1245-042&sd&13.60&7 (7x1)& 0.18 & 1.0\tabularnewline
PG 2151+100&sdB&12.68&9 (3x3)&0.39&1.0 \tabularnewline
EC 13047-3049&sdB&12.78&2 (1x2)& $<$0.10 & 1.0\tabularnewline
PG 1505+074&sdB&12.37&2 (2x1)&$<$0.10&1.0 \tabularnewline
LS IV -14 116&He-sdOB&12.98&2 (2x1)&0.11&1.0\tabularnewline
EC 12578-2107&sdB&13.52&7 (7x1)& 0.25 &1.0\tabularnewline
EC 13080-1508&sdB&13.65&3 (3x1)& 0.18 &1.0\tabularnewline
PB 8783&sdO+F&12.23&6 (3x2)&0.25&1.1 \tabularnewline
MCT 2341-3443&sdB&10.92&4 (2x2)&0.18&1.1  \tabularnewline 
EC 21595-1747&sdOB&12.62&4 (2x2)&0.18&1.1 \tabularnewline
PG 1230+067&He-sdOB&13.12&2 (1x2)&0.11&1.1 \tabularnewline
EC 15103-1557&sdB&12.82&6 (3x2)&0.25&1.1  \tabularnewline
PG 2313-021&sdB&13.00&6 (3x2)&0.18&1.1 \tabularnewline
PG 2349+002& sdB & 13.27 & 10 (1x10) & 0.32 & 1.1 \tabularnewline
PG 1207-033&sdB&13.34&2 (2x1)& 0.11 &1.1\tabularnewline
PG 1303-114&sdB&13.63&5 (5x1)& 0.18 &1.1\tabularnewline
PG 1343-102&sdB&13.69&4 (4x1)&0.25&1.1\tabularnewline
{\textbf{Suspended}}&&&&&\tabularnewline
LS IV+06 5&sdB&12.37&8 (4x2)&-&$>$1.3\tabularnewline
EC 14338-1445&sdB&13.55&2 (2x1)&-&$>$1.5\tabularnewline
EC 14599-2047&sdB&13.57&3 (3x1)&-&$>$1.5\tabularnewline
EC 01541-1409&sdB&12.27&12 (4x3)&-&$>$1.5\tabularnewline 
TYC 1077-218-1&sdOB&12.41&3 (3x1)&-&$>$2.0\tabularnewline 
LS IV +09 2&sdB&12.69&4 (2x2)&-&$>$2.0\tabularnewline
TYC 467-3836-1&sdB&11.70&6 (6x1)&-&$>$2.0\tabularnewline
\hline\hline
\end{tabular}
\end{center}
\end{table*}

\begin{figure*}[!ht]
\centering
\includegraphics[scale=0.45,angle=0]{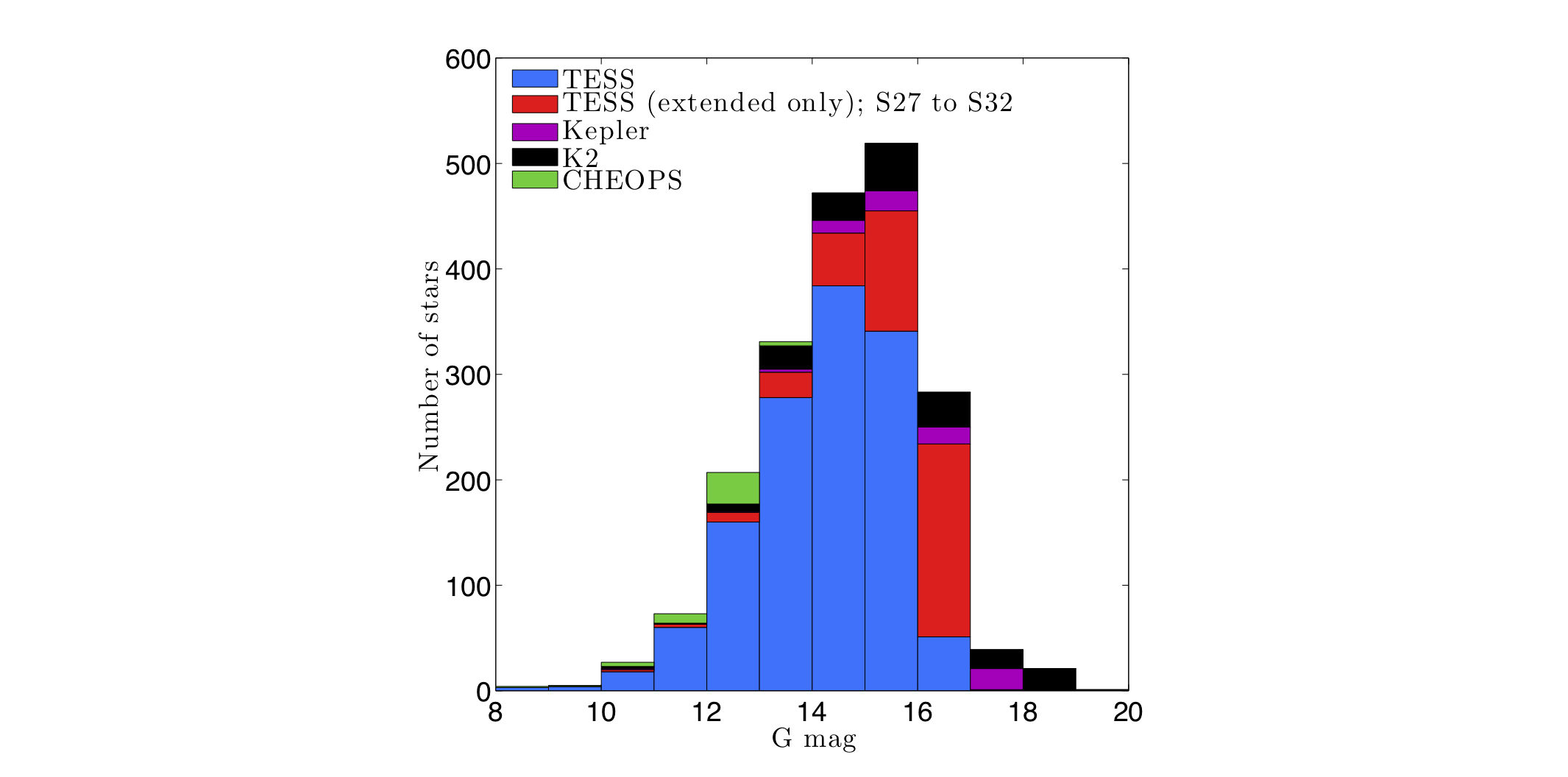}\\
\includegraphics[scale=0.1,angle=0]{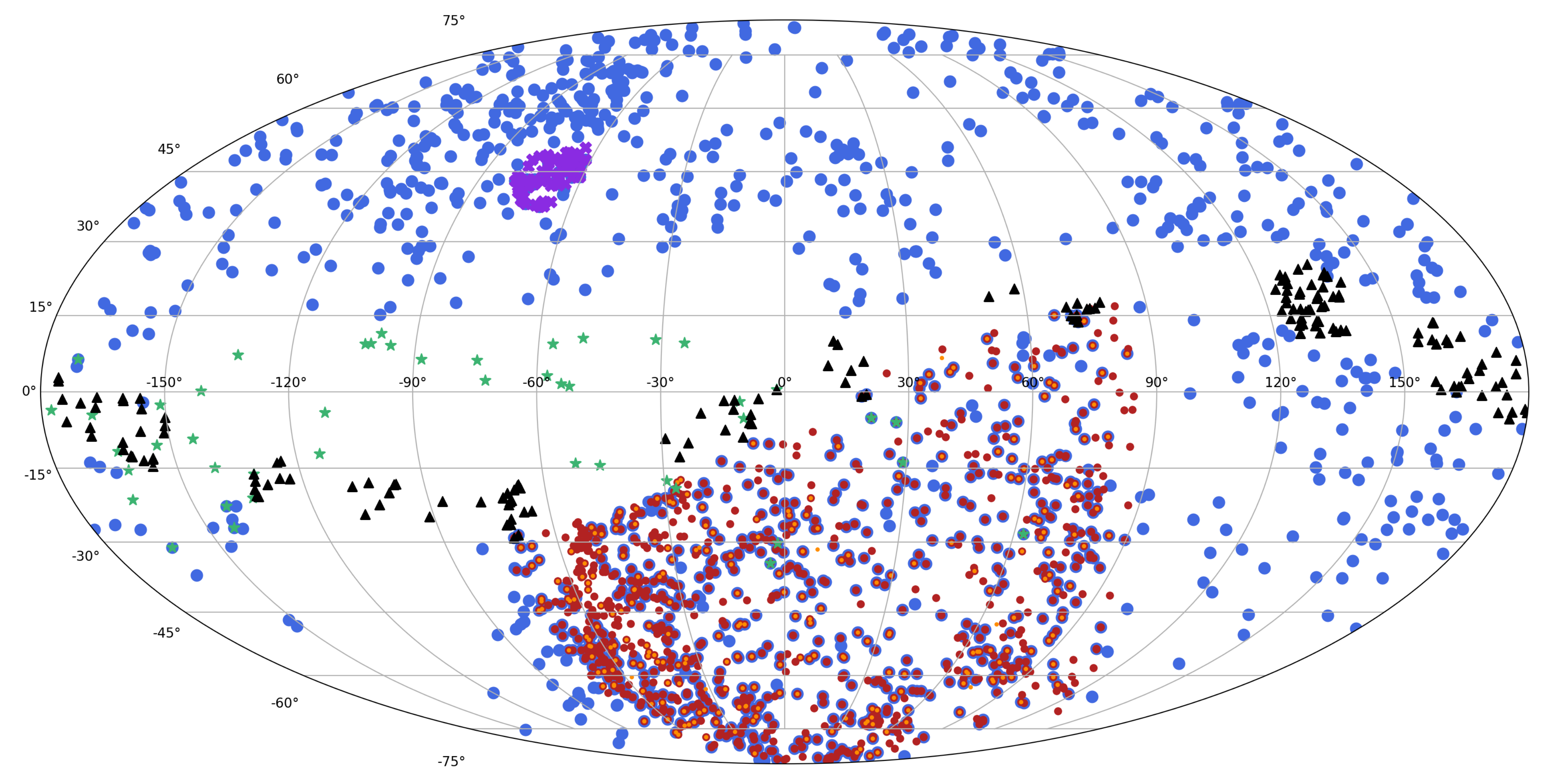}
\caption{\label{histo_alltargets} \textit{Top}: Number of hot subdwarfs per G magnitude bin observed by Kepler, K2, TESS (hot subdwarfs observed in primary mission, which are almost all reobserved in the extended mission), TESS extended only (hot subdwarfs observed for the first time in the extended mission; S27 to S32), and CHEOPS (as of December 19, 2020). \textit{Bottom}: Celestial distribution of these hot subdwarfs: TESS primary mission (blue dots), TESS extended mission 2-min and 20 s (red and dark orange dots), Kepler (purple crosses), K2 (black triangles), and CHEOPS (green stars).}  
\end{figure*}

CHEOPS (CHaracterising ExOPlanets Satellite) is a European Space Agency (ESA) mission primarily dedicated to the study of known extrasolar planets orbiting bright (6$<$V$<$12) stars. It was successfully launched into a 700 km altitude Sun-synchronous 99-min orbit on 18 December 2019. CHEOPS is a 30 cm (effective) aperture telescope optimised to obtain high-cadence high-precision photometric observations for a single star at a time in a broad optical band. CHEOPS is a pointed mission with mostly time-critical observations, so it has $\sim$20\% of free orbits that are partly used to observe bright, apparently single hot subdwarfs as a "filler" program (program ID002). This means that hot subdwarf observations are carried out when CHEOPS has no time-constrained or higher priority observations. The selected targets are generally close to the ecliptic where CHEOPS has its maximum visibility (but were not observed by K2) and are observed for one to three consecutive orbits, for a goal of a total of 18 orbits per target per season. As of December 19, 2020 (8 months after starting the program), 46 hot subdwarf targets have been observed by CHEOPS for a total of 290 orbits. The exposure time is 60 sec for all these targets (except HD 149382, for which it is 41 s). The list of CHEOPS targets and details can be found in Table \ref{CHEOPS_targets} ("Phase coverage" and "Minimum planet size" columns will be explained in Sect. \ref{Perf_CHEOPS}), as well as on {\url{https://github.com/franpoz/Hot-Subdwarfs-Catalogues}}. The nominal duration of the CHEOPS mission is 3.5 years (i.e., end 2023), when we hope to reach a total of about 25-30 orbits on 50-60 targets.

Finally, for completeness, let us mention that the CoRoT satellite \citep{2006cosp...36.3749B} performed for $\sim$24 days a high quality, nearly continuous photometric observation of the sdB pulsator KPD 0629-0016 \citep{2010A&A...516L...6C}. We will add these observations in our transit survey, but we will not carry out here a performance test for this one star.

Figure \ref{histo_alltargets} (top) summarizes, ranked per bin of G magnitudes, the available sample of hot subdwarf space-based light curves, as obtained from Kepler, K2, TESS (primary mission, as well as hot subdwarfs observed for the first time in the extended mission), and CHEOPS (as of December 19, 2020). No G magnitude is available for a few Kepler and K2 targets, so they have been included in Fig. \ref{histo_alltargets} with Kp minus 0.1, which is the mean difference between Kp and G magnitudes observed for targets for which both estimates are available. Figure \ref{histo_alltargets} (bottom) shows the celestial distribution of this sample.  

\subsection{Tools for transit searches in space-based light curves}
\label{sherlock}
To search for transit events we will make use of our custom pipeline {\fontfamily{pcr}\selectfont SHERLOCK} \citep{pozuelos2020a}\footnote{{The \fontfamily{pcr}\selectfont  SHERLOCK} (\textbf{S}earching for \textbf{H}ints of \textbf{E}xoplanets f\textbf{R}om \textbf{L}ightcurves \textbf{O}f spa\textbf{C}e-based see\textbf{K}ers)
code is fully available on GitHub site: \url{https://github.com/franpoz/SHERLOCK}}. This pipeline provides the user with easy access to Kepler, K2 and TESS data, for both SC and LC. The pipeline searches for and downloads the Pre-search Data Conditioning Simple Aperture (PDC-SAP) flux data from NASA's Mikulski Archive for Space Telescope (MAST). Then, it uses a multi-detrend approach via the  \texttt{WOTAN} package \citep{Hippke2019Python}, whereby the nominal PDC-SAP flux light curve is
detrended a number of times using a bi-weight filter or a Gaussian process, by varying the window size or the kernel size, respectively. This multi-detrend approach is motivated due to the associated risk of removing transit signals, in particular short and shallow ones. Then, each of the new, detrended light curves, jointly with the nominal PDC-SAP flux, are processed through the {\sc{transit least squares}} package (TLS) \citep{Hippke2019TransitPlanets} in the search for transits. Contrary to the classical Box Least Square (BLS) algorithm \citep{kovacs2002}, the TLS algorithm uses an analytical transit model that takes the stellar parameters into account. Then, it phase folds the light curves over a range of trial periods ($P$), transit epochs ($T_{0}$), and transit durations ($d$). It then computes $\chi^{2}$ between the model and the observed values, searching for the minimum $\chi^{2}$ value in the 3D-parameter space ($P$, $T_{0}$ and $d$). It is found that TLS is more reliable than classical BLS in finding any kind of transiting planet, and it is particularly suited for the detection of small planets in long time series, such as these coming from Kepler, K2 and TESS. TLS also allows the user to easily fine-tune the parameters to optimize the search in each case, which is particularly interesting when one deals with shallow transits. In addition, {\fontfamily{pcr}\selectfont  SHERLOCK}
incorporates a vetting module combining \texttt{TPFplotter} \citep{tpfplotter2020}, \texttt{LATTE} \citep{latte2020}, and \texttt{TRICERATOPS} \citep{giacalone2021} packages, which allows the user to explore any contamination sources in the photometric aperture used, momentum dumps, background flux variations, $x$--$y$ centroid positions, aperture size dependencies, flux in-and-out transits, each individual pixel of the target pixel file, and estimate the probabilities for different astrophysical scenarios such as transiting planet, eclipsing binary, eclipsing binary with twice the orbital period, among others. Collectively, these analyses help the user estimate the reliability of a given detection. 

For each event that overcomes the vetting process, the user may wish to perform ground- or space-based follow-up observations
to confirm the transit event on the target star, which is particularly critical for TESS observations due to the large pixel size (21~arcsec) and point spread function (which could be 
as large as 1~arcmin). These aspects increase the probability of contamination by a nearby eclipsing binary \cite[see e.g.][]{gunther2019,kostov2019,quinn2019,nowak2020}. However, the results coming directly from the searches performed with {\fontfamily{pcr}\selectfont SHERLOCK} via the TLS algorithm are not optimal; that is, the associated uncertainties of $P$, $T_{0}$ and $d$ are large, and their temporal propagation make their use to compute future observational windows and schedule a follow-up campaign impractical. Hence, {\fontfamily{pcr}\selectfont SHERLOCK} uses the results coming from TLS as priors to perform model fitting, injecting them into \texttt{allesfitter} \citep{allesfitter-code,allesfitter-paper}. The user can then choose between Nasted Sampling or MCMC analysis, whose posterior distributions are much more refined, with significant reductions of a few order of magnitudes of the uncertainties of $P$, $T_{0}$ and $d$. This allows us to schedule a follow-up campaign where the observational windows are more reliable.

\section{Injection-and-recovery tests}
\label{inj}
\begin{figure*}[!ht]
\begin{center}
\begin{tabular}{lll}
\includegraphics[scale=0.6,angle=0]{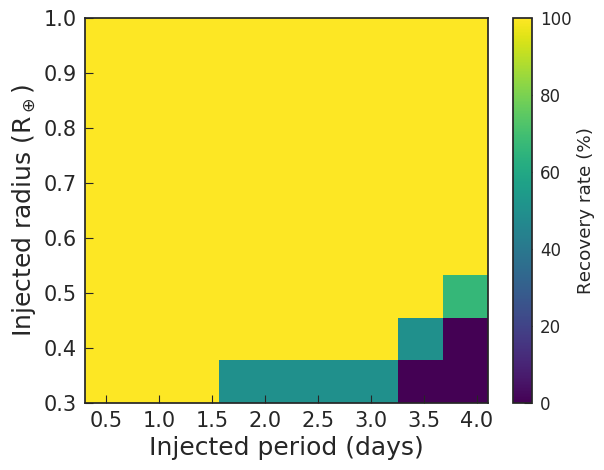}
\includegraphics[scale=0.6,angle=0]{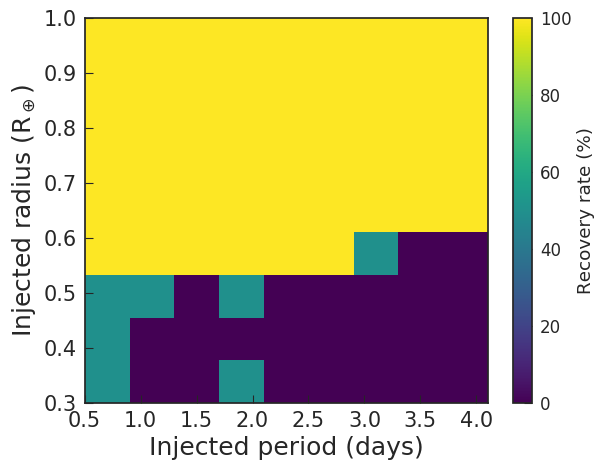}
\end{tabular}
\end{center}
\caption{\label{InjandrecKepler} Injection-and-recovery tests. \textit{Left panel}: KIC 8054179 (Kp=14.40, G=14.34), based on Q6 Kepler data (90 days). \textit{Right panel:} EPIC 206535752 (Kp=13.99, G=14.10), observed during Campaign 3 of K2 (81 days). Injected transits of planets have 0.3-1.0 $R_{\Earth}$ (steps of 0.1~$R_{\Earth}$) with 0.5-4.1~d (steps of 0.2~d) orbital periods.}  
\end{figure*}

To quantify the detectability of transiting bodies in our sample of hot subdwarfs, we performed a suite of injection-and-recovery tests. While the detectability of a transit is a target-and-sector/quarter dependent issue, these experiments allow us to verify the general reliability of our survey. We explored several data sets coming from the Kepler, K2 and TESS missions, and for each one we chose a range of stellar magnitudes to study. In all cases, we followed the procedure described by \cite{pozuelos2020a} and \cite{demory2020}; that is, we downloaded the PDC-SAP fluxes in each case and generated a grid of synthetic transiting planets by varying their orbital periods and radii, which are injected in the downloaded light curves. We then detrended the light curves and search for the injected planets. The search itself is done by applying the simple TLS algorithm. Indeed, the multi-detrend approach applied by {\fontfamily{pcr}\selectfont SHERLOCK} makes it more efficient at finding shallow-periodic transits, but with a higher computational cost. Hence, the full use of {\fontfamily{pcr}\selectfont SHERLOCK} in the injection-and-recovery experiments is too expensive. That means that our findings in these experiments might be considered as upper limits for the minimum planet sizes, and during our survey we might detect even smaller planets. We defined a synthetic planet as ``recovered'' when we detect its epoch with one-hour accuracy and if we find its period better than 5$\%$. Depending in the number of sectors/quarters available, we explored the $R_{\mathrm{planet}}$--$P_{\mathrm{planet}}$ parameter space in different ranges. We conducted two different experiments to qualify the performances achievable with Kepler, K2 and TESS data. The first experiment consisted in full injection-and-recovery tests focusing on a region of the parameter space corresponding to small close-in exoplanets. For Kepler and K2, the injected planets range of 0.3 to 1.0~$R_{\Earth}$ with steps of 0.1~$R_{\Earth}$, and 0.5--4.1~d with steps of 0.2~d, for a total of 152 scenarios. For TESS, the injected planets have 0.5--3.0~$R_{\Earth}$ with steps of 0.05~$R_{\Earth}$, and 1.0--6.0~d with steps of 0.1~d, hence for a total of 2500 scenarios for each of them. However the 6-sector test, due to computational cost, was done with 0.5--2.5~$R_{\Earth}$ with steps of 0.2~$R_{\Earth}$, and 1.0--5.0~d with steps of 0.2~d, for a total of 200 scenarios (as a corollary, injection-and-recovery tests on more sectors, up to 13 sectors for one-year-continuous observations, are out of our reach). The second experiment concerned farther and bigger planets, that is, up to 10.0~$R_{\Earth}$ and 35 d orbital period. Full injection-and-recovery tests, with sufficiently small steps, led to a too important computational cost. Instead, we chose to focus on particular periods (1, 5, 15, 25 and 35~d), and to determine for these periods what is the minimum planet size detectable for each set of data considered. This limit of 35~d is justified by the transit probability, which quickly falls to very low values with increasing orbital period (for a typical hot subdwarf of 0.15 $R_{\odot}$ and 0.5 $M_{\odot}$, the transit probabilities at 10~d and 50~d are about 1\% and 0.35\% respectively). For each period we explored $\sim$30 scenarios with fine steps of 0.1~d and 0.1~$R_{\Earth}$, around a nominal value of the size previously computed with an exploration of sizes from 1 to 10~$R_{\Earth}$ with steps of 1~$R_{\Earth}$. This strategy allows us to obtain robust estimations of the sizes with a recovery rate $\gtrsim$ 90\% for each period explored, with a total number of scenarios considerably higher than for full injection-and-recovery maps as in Fig.~\ref{InjandrecKepler}$-$\ref{InjandrecTESS_multi}. The results coming from these experiments are consequently better grounded than those coming from the full maps, which might be considered as more rough estimations of the recovery rates, but with a better and quicker general overview for small and close-in exoplanets. We of course do not exclude to find longer period planets during our transit survey. But for the tests carried out here whose purpose is to quantify our potential to find planets by a transit survey, the limit of 35~d is a good balance between the computational cost and the probability of transit. For Kepler data we explored one quarter, which corresponds to $\sim$~90 d of data, as well as monthly subquarters. A similar approach is adopted for K2 data, where observations usually span one campaign of $\sim$~80 d of data and sub-samples of 30 d. Finally, for TESS data, we tested data covering 1, 2, 3 and 6 sectors (27 to 162 d). 

In all experiments, we assumed that the host star is a canonical hot subdwarf with a radius of $0.175 \pm 0.025~R_{\odot}$ and a mass of $0.47\pm 0.03~M_{\odot}$. We considered only SC data here. We selected targets that are as 'unremarkable' as possible, being non-variable stars (i.e., no peak emerges above 4$\sigma$ in a Lomb-Scargle periodogram, from pulsations, from reflection/ellipsoidal effect due to a binary nature, or from any other kind of variability) and exhibiting quiet (low scatter) light curves. Indeed, the experiments performed here, based on injection of synthetic transits, apply for computational cost reasons one detrending only to the resulting light curves. In all cases we used the bi-weight method with a nominal window-size of 2.5~h, large enough to protect short transits of close-in exoplanets (which have a typical duration of $\sim$20 min) and to remove most of the stellar noise, variability and instrumental drifts. For the actual transit search, the light curves will be detrended twelve times using either a bi-weight filter or a Gaussian process (Sect. \ref{sherlock}), which allows us to optimize the planet search and increasing the detectability of small planets. In this context, it is therefore important for our injection-and-recovery experiments here to select targets as quiet as possible (minimizing the need of detrending), in order to obtain results as representative as possible.

\subsection{Results for Kepler and K2}
\begin{table*}[!ht]
\caption{\label{tableinjects}Minimum size of planets in units of $R_{\Earth}$ detectable in typical light curves with a $\gtrsim$ 90\% recovery rate. All stars have $0.175 \pm 0.025~R_{\odot}$ and $0.47\pm 0.03~M_{\odot}$.}
\begin{center}
\begin{tabular}{lccccccc}
\hline\hline
\noalign{\smallskip}
\multirow{2}{*}{Object ID} & \multirow{2}{*}{G Mag} & Data &  \multirow{2}{*}{1~d} &  \multirow{2}{*}{5~d} &  \multirow{2}{*}{15~d} &  \multirow{2}{*}{25~d} &  \multirow{2}{*}{35~d} 
\tabularnewline
 & & length (d)  & &&&& 
\tabularnewline
\hline
\noalign{\smallskip}
\textit{Kepler} & & & & & & \tabularnewline
8054179 & 14.3 & 90 & 0.3 & 0.5 & 0.8 & 1.0 & 1.2 \tabularnewline
        & & 30 & 0.5 & 0.6 & 1.0 & -- & -- \tabularnewline
3353239 & 15.2 & 30 & 0.6 & 0.8 & 1.1 & -- & -- \tabularnewline
5938349 & 16.1 & 30 & 0.7 & 1.1 & 2.0 & -- & -- \tabularnewline
8889318 & 17.2 & 30 & 0.9 & 1.2 & 2.4 & -- & -- \tabularnewline
5342213 & 17.7 & 30 & 1.2 & 1.7 & 3.2 & -- & -- \tabularnewline 
\hline
\noalign{\smallskip}
\textit{K2} & & & & & & \tabularnewline
206535752 & 14.1 & 80 & 0.6 & 0.8 & 1.0 & 1.5 & 2.1 \tabularnewline 
          & & 30 & 0.6 & 0.9 & 1.6 & -- & -- \tabularnewline 
211421561 & 14.9 & 30 & 0.7 & 1.4 & 1.9 & -- & --  \tabularnewline 
228682488 & 16.0 & 30 & 1.0 & 1.4 & 2.5 & -- & -- \tabularnewline 
251457058 & 17.1 & 30 & 1.4 & 2.3 & 3.4 & -- & -- \tabularnewline
248840987 & 18.1 & 30 & 2.1 & 3.3 & 5.4 & -- & -- \tabularnewline
\hline
\noalign{\smallskip}
\textit{TESS} & & & & & & \tabularnewline
147283842 & 10.1  & 27 & 0.5 & 0.7 & 1.5 & -- & -- \tabularnewline 
362103375 & 13.0 & 27 & 1.0 & 1.7 & 2.0 & -- & -- \tabularnewline
 & & 162 & 0.7 & 0.8 & 0.9 & 1.0 & 1.3 \tabularnewline
096949372 & 13.0 & 27 & 1.1 & 1.8 & 2.0 & -- & -- \tabularnewline
441713413 & 13.1 & 27 & 1.3 & 1.7 & 2.0 & -- & -- \tabularnewline
 &  & 54 & 1.3 & 1.7 & 1.9 & >10 & >10 \tabularnewline
085400193 & 14.1 & 27 & 1.8 & 2.3 & 2.8 & -- & -- \tabularnewline
220513363 & 14.1 & 27 & 1.6 & 1.8 & 2.7 & -- & -- \tabularnewline
 & & 81 & 1.3 & 1.6 & 2.5 & 3.0 & 3.0 \tabularnewline
000008842 & 15.0 & 27 & 2.7 & 3.2 & 4.7 & -- & -- \tabularnewline

\hline\hline
\end{tabular}
\end{center}
\end{table*}

Figure~\ref{InjandrecKepler} (left) shows the full injection-and-recovery test for KIC 8054179 (Kp=14.40, G=14.34) from the Q6 Kepler data (90 days). We found that planets smaller than $\sim$0.4~$R_{\Earth}$ with orbital periods larger than $\sim$1.5 days, and smaller than $\sim$0.5~$R_{\Earth}$ with orbital periods larger than $\sim$3.2 days, have recovery rates below 50\%, i.e., we will most likely be unable to detect them (Fig.~\ref{InjandrecKepler}). For the shortest orbital periods ($\lesssim$ 1.5 d), objects as small as $\sim 0.3~ R_{\Earth}$ are fully recovered\footnote{we explicitly checked that the detection rate of planets below $\sim 0.3~ R_{\Earth}$ quickly falls below 50\%.}. Results from the second experiment focusing on farther and larger planets are presented in Table \ref{tableinjects}: the smallest planet detectable for 1~d to 35~d increases from 0.3 to 1.2 $R_{\Earth}$ for KIC 8054179, considering 90~d of data. 

We also performed similar experiments for 4 other representative Kepler targets with increasing magnitudes, for one subquarter, i.e., 1 month of data (we also provided results for 1 month data for KIC 8054179, for comparison purposes). Results are presented in Table \ref{tableinjects}. For a typical Kepler target of 16th G magnitude (see Fig. \ref{histo_alltargets}), a sub-Earth size planet of 0.7 $R_{\Earth}$ can still be detected at 1~d period, and a 2.0 $R_{\Earth}$ at 15~d period (considering one month data). 

Figure \ref{InjandrecKepler} (right) shows the full injection-and-recovery test for EPIC~206535752 (Kp=13.99, G=14.10), observed during Campaign 3 of K2 (81 days). We find that $\sim$0.6~$R_{\Earth}$ planets are fully recovered up to $\sim 3$ d orbital periods, while the detectability of objects smaller than 0.5~$R_{\Earth}$ quickly drops below 50\% for all orbital periods, meaning that we will likely not be able to detect them. Results from the second experiment on EPIC~206535752 focusing on farther and larger planets are presented in Table \ref{tableinjects}: the smallest planet detectable for 1~d to 35~d quickly increases from 0.6 to 2.1~$R_{\Earth}$, considering 80~d of data. 

Similar experiments were also carried out for 4 other K2 targets with increasing magnitudes, with a sub-sample of 30 d data. Results are presented in Table \ref{tableinjects}. For a typical K2 target of 15th G magnitude (see Fig. \ref{histo_alltargets}), a sub-Earth size planet of 0.7~$R_{\Earth}$ can still be detected at 1~d period, as can a 1.9~$R_{\Earth}$ at 15~d period (considering one month of data). 

On a concluding remark, for a given magnitude and data duration, the Kepler performances are significantly superior to the K2 ones, although the K2 targets are generally brighter (Fig. \ref{histo_alltargets} and Table \ref{tableinjects}). Almost all Kepler targets will allow us to detect transiting objects with a radius $\lesssim$ 1.0 $R_{\Earth}$, while that is the case for about 2/3 of K2 targets.

\subsection{Results for TESS}

\begin{figure*}[!ht]
\begin{center}
\begin{tabular}{lll}
\includegraphics[scale=0.55,angle=0]{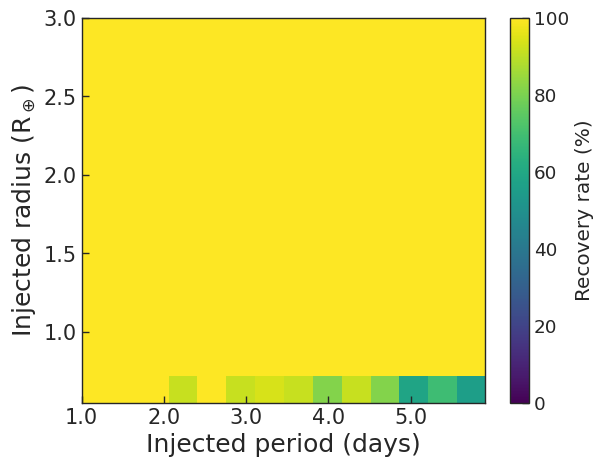}
\includegraphics[scale=0.55,angle=0]{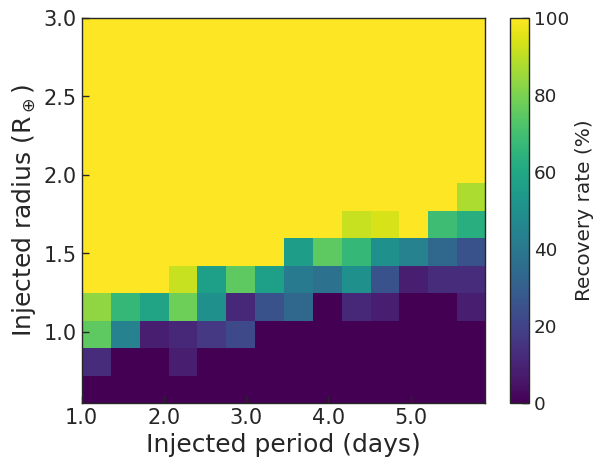}\\
\includegraphics[scale=0.55,angle=0]{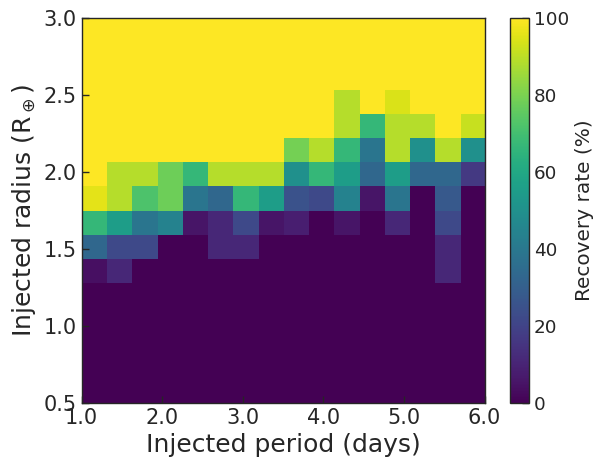}
\includegraphics[scale=0.55,angle=0]{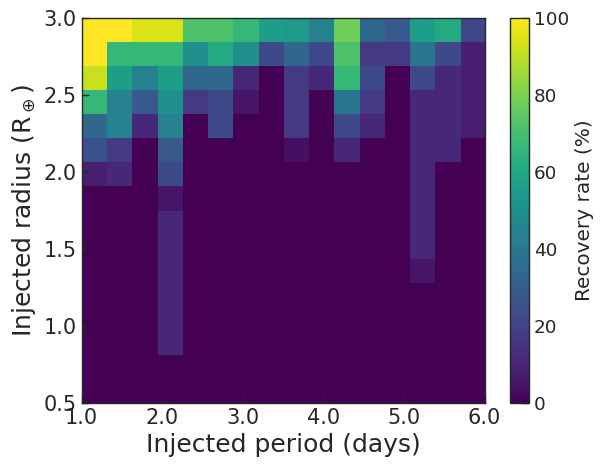}
\end{tabular}
\end{center}
\caption{\label{InjandrecTESS} Results of injection-and-recovery tests on 4 sdB stars observed for one sector by TESS: TIC 147283842 (G=10.1, top left panel), TIC 96949372 (G=13.0, top right panel), TIC 85400193 (G=14.1, bottom left panel), and TIC 000008842 (G=15.0, bottom right panel). 2500 injection-and-recovery tests were made for each star.
}  
\end{figure*}

Figure \ref{InjandrecTESS} presents results of injection-and-recovery tests for 4 stars observed in 1 sector by TESS. The four selected stars also are very quiet, non-variable stars. They have magnitudes of G=10.1, G=13.0, G=14.1, and G=15.0. Figure \ref{InjandrecTESS} shows that typically, $\sim$0.5 $R_{\Earth}$ (G$\sim$10.0), $\sim$1.2 $R_{\Earth}$ (G$\sim$13.0), $\sim$1.9 $R_{\Earth}$ (G$\sim$14.1), and $\sim$2.7 $R_{\Earth}$ (G$\sim$15.0) planets can be retrieved from TESS 1-sector light curves for the shortest orbital periods with a $\gtrsim$ 90\% recovery rate. 

\begin{figure*}[!ht]
\begin{center}
\begin{tabular}{lll}
\includegraphics[scale=0.55,angle=0]{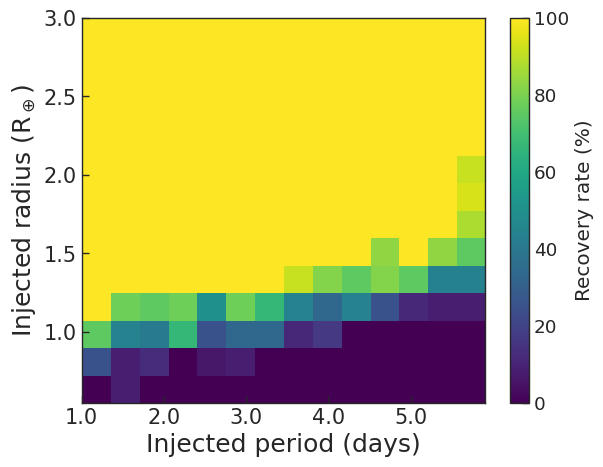}
\includegraphics[scale=0.55,angle=0]{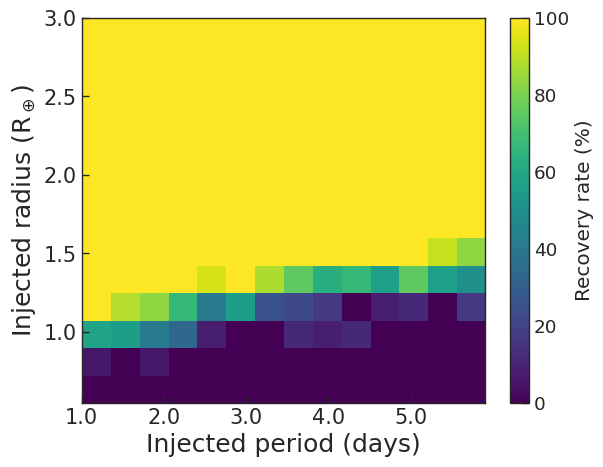}\\
\includegraphics[scale=0.55,angle=0]{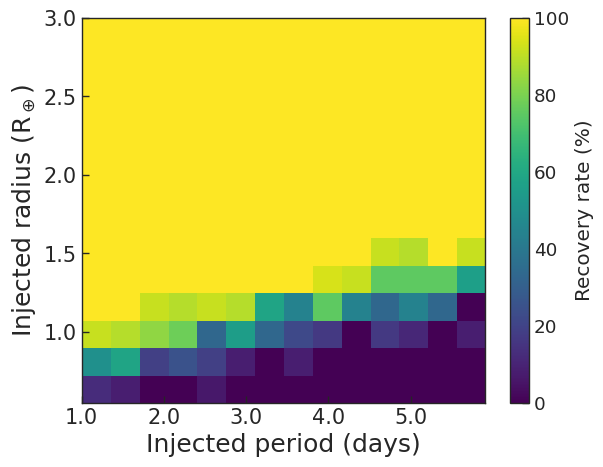}
\includegraphics[scale=0.55,angle=0]{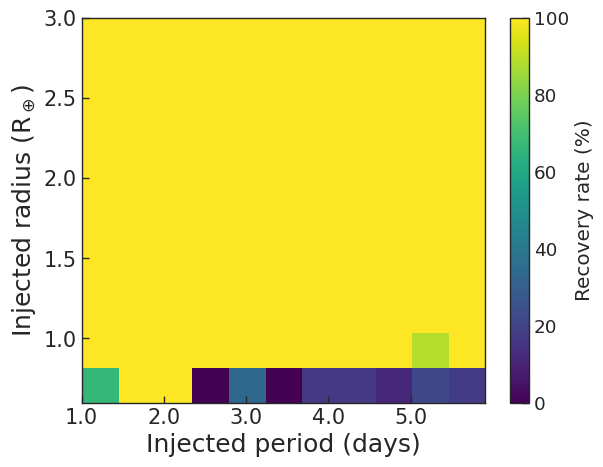}
\end{tabular}
\end{center}
\caption{\label{InjandrecTESS_multi} Results of injection-and-recovery tests on stars observed in multiple sectors by TESS. \textit{Top panels}: TIC 441713413 (G=13.07), 1-sector data (left) and 2-sector data (right). \textit{Bottom panels}: TIC 362103375 (G=13.04), 1-sector data (left) and 6-sector data (right).}  
\end{figure*}

To appreciate the increase of the detectability with multi-sector observations, we performed similar tests on TIC 441713413 (G=13.07) observed in 2 sectors (S16, S23), on TIC 220513363 (G=14.1) observed in 3 sectors (S1, S2, S3), and on TIC 362103375 (G=13.04) observed in 6 sectors (S14, S15, S18, S22, S25, and S26). All stars are compared to results from 1-sector-only tests (S16 for TIC 441713413, S1 for TIC 220513363, and S14 for TIC 362103375). Figure \ref{InjandrecTESS_multi} and Table 3 present and compare the results of these experiments. The improvement in detectability from 1 to 2 sectors is barely perceptible, and is noticed only for orbital periods beyond 5~d. This is an important result since the majority of TESS targets were observed for 1-sector only during the primary mission (Table \ref{sd_TESS_primary}), and will be reobserved for another 1-sector in the extended mission. No significant improvement of the detectability obtained from 1-sector primary mission (Fig. \ref{InjandrecTESS}) is therefore expected with 1 more sector data in the extended mission. The improvement from 1 to 3 sectors (TIC 220513363, see Table \ref{tableinjects}) and 6 sectors (TIC 362103375, see Fig. \ref{InjandrecTESS_multi} and Table \ref{tableinjects}) is increasingly noticeable: for example, we are now able to reach sub-Earth sized objects up to 25~d with 6 sectors, while it was only possible for orbital period of 1~d (and below) with 1 sector only.

A few more words can be said about the impact of the data length on the minimum detectable radius. In an ideal case, the longer the data set, the smaller the planet that can be detected thanks to the increased number of stacked transits, which improves the statistics and increases the signal-to-noise ratio (SNR). This is directly related to the working procedure of our transit search algorithm (TLS, see Sect. \ref{sherlock}). However, the real nature of light curves, which always present a level of noise that is impossible to remove, makes that we do not always have a clear improvement when staking more transits. This is in particular the case for the short orbital periods, adding more transits does not always yield a vast improvement, providing there is already a large number of them. This can be noticed in Table \ref{tableinjects} for orbital periods of 1~d, for e.g. KIC 8054179, EPIC 206535752, TIC 441713413. For longer orbital periods the effect is stronger, because the increase in the number of stacked transits is relatively more important. For example for the TESS sample, the improvement in the minimum size of planets detectable with an orbital period of 15~d is remarkable when we expand our analysis from 1 sector to 2 (TIC 441713413), 3 (TIC 220513363), and 6 sectors (TIC 362103375).

To finish this section, let us mention that Fig.~\ref{InjandrecKepler}$-$\ref{InjandrecTESS_multi} as well as Table \ref{tableinjects} also allows us to assess the general reliability of our results for Kepler, K2 and TESS light curves. While the detectability will be (unavoidably) target- (for similar magnitude) and/or sector/quarter- (for similar data length) dependent (and also because of the actual radius of the host star, of course), the general comparison of the tests carried out here gives consistent trends. For example, the results on three different stars with G$\sim$13.0  (TIC 096949372, 362103375 and 441713413) for 1-sector TESS data of three different sectors are globally consistent. 

\section{CHEOPS performances on hot subdwarfs}
\label{Perf_CHEOPS}

\begin{figure*}[!ht]
\begin{center}
\begin{tabular}{lll}
\includegraphics[width=0.99\textwidth]{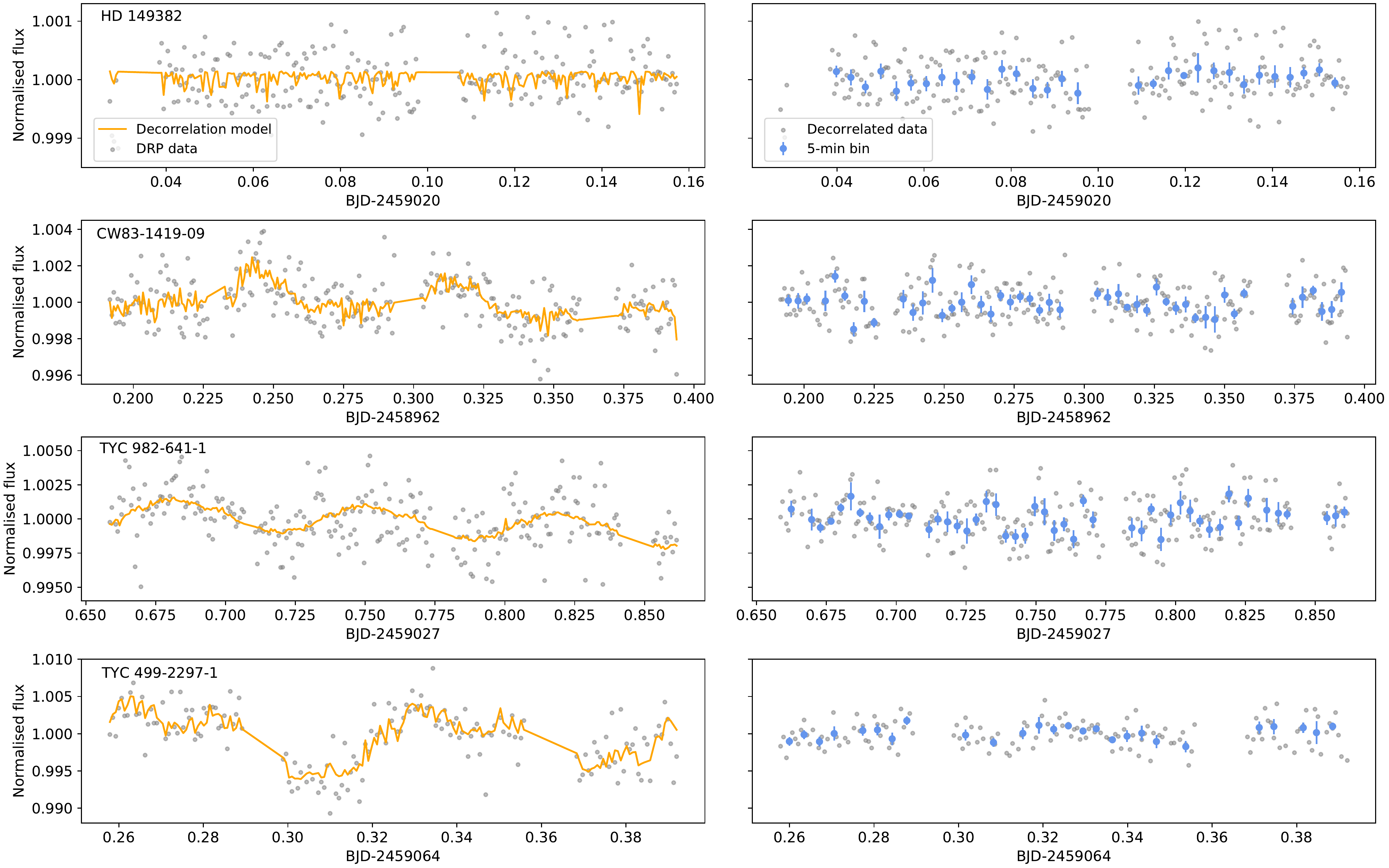}
\end{tabular}
\end{center}
\caption{\label{lc_CHEOPS} Representative light curves of hot subdwarfs produced by CHEOPS. 
From top to bottom: HD 149382 (G=8.9) in its fifth visit, CW83 1419-09 (G=12.0) in its first visit,
TYC 982-6141 (G=12.2) in its first visit, and TYC 499-2297-1 (G=12.6) in its fourth visit. In all cases, the raw light curves
as processed by the DRP (grey dots) are displayed in the left panels, jointly with the best decorrelation model (orange line) found 
by means of the {\tt{pycheops}} package. In the right panels, the decorrelated data (grey dots) with a 5-min bin (blue dots) are shown. The $y$-scale is the same for each pair of right-and-left panels.}  
\end{figure*}
\begin{figure*}[!ht]
\begin{center}
\begin{tabular}{lll}
\includegraphics[width=0.99\textwidth]{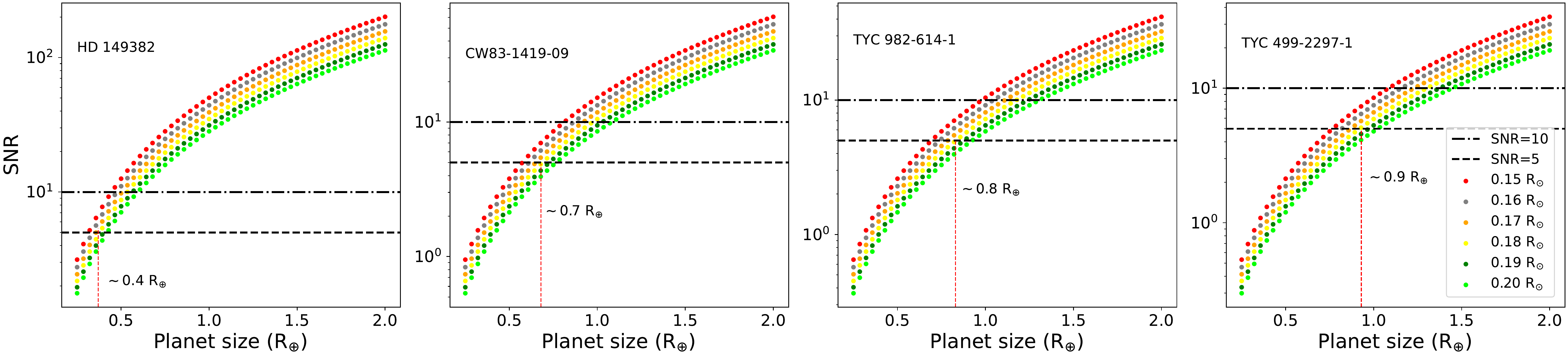}
\end{tabular}
\end{center}
\caption{\label{CHEOPS_1} Performances of CHEOPS on hot subdwarfs, assuming a single 20-min transit. From left to right panel: 
HD 149382 (G=8.9), CW83-1419-09 (G=12.0), TYC 982-614-1 (G=12.2), and TYC 499-2297-1 (G=12.6). The minimum planet size for SNR=5 and a 0.18 $R_{\odot}$ host is indicated next to the red vertical line.}  
\end{figure*}

Figure~\ref{lc_CHEOPS} displays typical light curves obtained by CHEOPS for four representative targets: 
(1) HD 149382, one of the brightest known sdB stars (G=8.9) which was not observed by TESS, Kepler or K2; 
(2) CW83-1419-09 (G=12.0) and (3) TYC 982-614-1 (G=12.2), which represent typical CHEOPS targets in terms of
magnitude; and (4) TYC 499-2297-1, a fainter target of G=12.6, which exceeds CHEOPS' standard
specifications. The light curves were processed using version 12 of the Data Reduction Pipeline (DRP; \citealt{2020A&A...635A..24H}). 

These light curves were obtained with the aperture that offers the smallest root mean square of variation in count rates, which is generally the \texttt{DEFAULT} one (which have a radius of 25 arcsec). Then, we evaluated how the flux was correlated with different parameters such as the time, CHEOPS roll angle, $x$--$y$ centroids, background, and contamination. This inspection was done
with the {\tt{pycheops}}\footnote{\url{https://github.com/pmaxted/pycheops}} package (v0.9.6), which is 
developed specifically for the analysis of CHEOPS data. Hence, we decorrelated the light curves of any unwanted trends, by calculating the Bayesian Information Criteria (BIC) of each combination of trends under the assumption that the combination that induces the lowest BIC best describes any trends. If needed, we also removed outliers. Once the light curves were decorrelated, we visually inspected them in the search for potential transits.  

The hot subdwarf observations made by CHEOPS are fillers. This results in light curves spanning 1.5 to 5 hr (with gaps due to Earth occultations and/or passages through the South Atlantic Anomaly, but always with a minimum efficiency of 60\% along an orbit, and always less than 100\%) separated by several days for a given target. This makes impractical the application of injection-and-recovery tests as conducted in Section~\ref{inj}. Indeed, our injection-and-recovery experiments were done via the TLS transit search tool, which is useful for long time-series observations such as the ones coming from Kepler, K2 or TESS. The power of TLS-based searching relies on the stacking of many transits, which eventually increases the signal-to-noise ratio of a given periodic signal. However, the CHEOPS light curves are short observational data sets, in which we expect to find single transits. Hence, to characterize CHEOPS' performance on hot subdwarfs, we estimated the minimum planet size detectable based on the transit depth that could be detected with a SNR of 5, assuming a transit duration of 20 minutes (which is the typical duration for a $\sim$~12~h orbital period) and for various typical stellar radii from 0.15 to 0.20 $R_{\odot}$. 

The noise of the light curve is estimated by the {\tt{pycheops}} package using the scaled noise method. It assumes that the noise in the light curve is white noise with standard error $b$ times the error values provided by the CHEOPS DRP. We then inject transits into the light curve and find the transit depth such that the SNR of the transit depth measurement is 1. The transit model used for this noise estimate includes limb darkening so we define the depth as $D=k^2$ where $k$ is the planet-star radius ratio used to calculate the nominal model. We can use a factor $s$ to modify the transit depth in a nominal model ${\bf m_0}$ calculated with approximately the correct depth to produce a new model
 $ {\bf m}(s) = 1 + s\times({\bf m_0}- 1)$. If the data are normalised fluxes
${\bf f} = f_1, \dots, f_N$ with nominal errors ${\bf \sigma} =
\sigma_1,\dots, \sigma_N$ then the log-likelihood for the model given the data is
\[\ln {\cal L}  = -\frac{1}{2b^2}\chi^2 - \frac{1}{2}\sum_{i=1}^N \ln \sigma_i^2  - N\ln b - \frac{N}{2}\ln(2\pi)\]
where
$\chi^2 = \sum_i^N \left(f_i - 1 - s(m_{0,i}-1\right)^2/\sigma_i^2. $ The maximum likelihood occurs for parameter values $\hat{s}$, and $\hat{b}$ such that
$\left. \frac{\partial  \ln {\cal L}}{\partial s}\right|_{\hat{s},\hat{b}} = 0$
and
$\left. \frac{\partial  \ln {\cal L}}{\partial b}\right|_{\hat{s},\hat{b}} = 0$,
from which we obtain
 \[\hat{s} = \sum_{i=1}^N \frac{(f_i - 1)(m_{0,i}-1) }{\sigma_i^2} \left[ \sum_{i=1}^N \frac{(m_{0,i}-1)^2}{\sigma_i^2}\right]^{-1}\]
 and
 \[\hat{b} = \sqrt{\chi^2/N}.\] The standard errors on the eclipse depth if $s\approx 1$ is  \[\sigma_D = D b\left[\sum_{i=1}^N \frac{(m_i-1)^2}{\sigma_i^2}\right]^{-1/2}.\]

Figure~\ref{CHEOPS_1} shows the minimum planet sizes
detectable at SNR=5 for our four representative targets. For HD 149382, a $\sim$0.4~$R_{\Earth}$ object 
(for a 0.18~$R_{\odot}$ host) could be detected at SNR=5 if transiting. CW83-1419-09 and TYC 982-614-1 
exhibit typical results for CHEOPS targets, reaching detections of $\sim$0.7-0.8~$R_{\Earth}$ objects. 
Finally, the fainter TYC 499-2297-1 could allow the detection of a $\sim$0.9~$R_{\Earth}$ object. 
The minimum planet size for all CHEOPS targets can be found in Table~\ref{CHEOPS_targets}, which are given for  
a 0.18~$R_{\odot}$ host and for SNR=5 in all cases. 

\begin{figure}
\includegraphics[width=0.99\columnwidth]{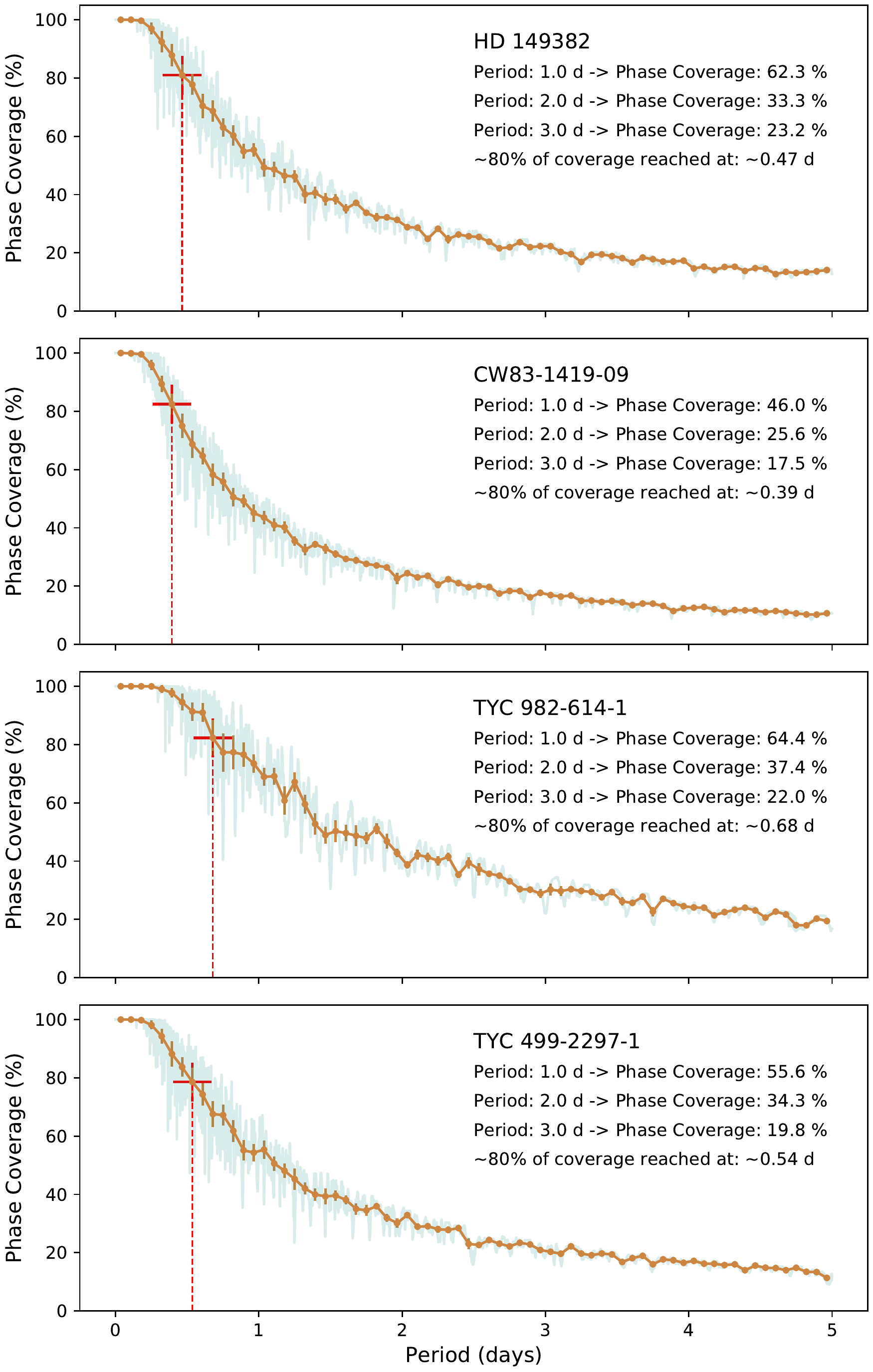}
\caption{\label{CHEOPS_2} Phase coverage (in \%) as a function of orbital period reached after one season of observations with CHEOPS. From top to bottom panel: HD 149382 (7x2 orbits), CW83 1419-09 (4x3 orbits), TYC 982-614-1 (6x3 orbits), and TYC 499-2297-1 (6x2 orbits). In all cases, the blue lines represent the full range of 5000 periods explored and the orange lines the binning each $\sim$1.7~hr. The orbital periods for which the phase coverages are $\sim$80\% are marked with dotted vertical red lines.}  
\end{figure}

Another important property to determine, given the filler nature of CHEOPS observations, is which 
orbital periods (and for which coverage of the orbit) are reached with the existing observations. This was measured by computing the phase coverage of an hypothetical planet in a range of periods. More precisely, we computed the 
percentage of the phase covered for each orbital period from $P_{\rm orb}=$0.001~d to 5~d, in intervals of 
0.001~d. Hence, we evaluated the phase coverage for a total of 5000 periods. Then, to aid
interpreting the phase coverage at different periods, we binned the periods by 1.7~hr. To illustrate the current status of our observational program, for each target we estimated the period at which a phase coverage of $\sim$80\% is reached, meaning that periods equal to or shorter than this would most likely be detected if the planet exists and transits. However, even if the probabilities are low, a hypothetical planet may still reside in the unexplored phase.
Results for our four representative targets are presented in Fig.~\ref{CHEOPS_2}.
As of 19 December 2020, for our four representative cases we found phase coverage of $\sim$80\% for
orbital periods of $\sim$0.47~d, $\sim$0.39~d, $\sim$~0.68 d, and $\sim$0.54~d for, respectively, HD 149384 (7x2 orbits), 
CW83 1419-09 (4x3 orbits), TYC 982-614-1 (6x3 orbits), and TYC 499-2297-1 (6x2 orbits). The orbital periods reached for a
phase coverage higher than 80\% for the CHEOPS targets can be found in Table \ref{CHEOPS_targets}. 

In light of the results of the injection-and-recovery tests in the Kepler, K2 and TESS light curves, all CHEOPS
targets with a minimum detectable planet size greater than $\gtrsim 1.1 R_{\Earth}$ have been suspended (see Table \ref{CHEOPS_targets}). These targets generally have fainter magnitudes, or are located in a crowded field, or have a bright close contaminating object, 
which explains the poorer ability to detect planets around these objects. Another explanation is that some targets are pressure-mode (p-mode) sdB pulsators with a relatively high amplitude (this is the case for EC 15041-1409 and TYC 1077-218-1), 
which are not properly removed with our current detrending procedure (this is an improvement we aim to implement
in the coming months). We instead chose to focus on the most promising targets having planets detectable below $\lesssim 1.1 R_{\Earth}$,
because in these cases, CHEOPS will notably contribute to increasing the number of targets for which we could detect planetary
remnants (which are likely small, possibly disintegrating objects) around post-RGB stars. From 
Tables~\ref{sd_TESS_primary}, \ref{sd_Kepler} and \ref{sd_K2}, and the results from Table~\ref{tableinjects}, it 
is estimated that about 160 stars observed by Kepler and K2 (almost all of them for Kepler, and about 2/3 of them for K2), and about 50 stars from TESS (the very brightest ones, and those with 
$G\lesssim 13.0$ observed for at least $\sim$ 6 sectors), will reach this minimum planet size. 
Statistically only $\sim$40\% of them are single hot subdwarfs, while in contrast, all CHEOPS 
targets have been chosen to be, to the best of our knowledge, single hot subdwarfs (or, in a few cases, subdwarfs in wide binary systems). 

The orbital periods reached by CHEOPS' filler observations will remain modest (about 1~d orbital period with a $80\%$ phase coverage by the end of the mission for most targets). However, these results are valuable for placing constraints on 
the survival rates of planets that are engulfed in the envelope of their red giant host. Such remnants, if present, are expected to 
have very short orbital periods due to the orbital decay of orbit of the inspiraling planet inside its host star. It is noteworthy here that all of the five Earth-sized planets suspected around KIC 05807616 and KIC 10001893 have orbital periods of a 
few hours only; \citealt{2011Natur.480..496C,2014A&A...570A.130S}, and all known sdB+red dwarf or brown dwarf post-CE binaries have orbital periods below 1~d \citep{2018A&A...614A..77S,2019A&A...630A..80S,2021MNRAS.501.3847S}. Finally, CHEOPS provides an excellent opportunity 
to observe very promising targets, such as HD 149382, which have not been observed by Kepler, K2 or TESS.

\section{Conclusions and future work}
\label{cc}
This paper presented our project to perform a transit survey to search for planets around hot subdwarfs. While no such planetary transit have been found to date, high-quality photometric light curves are now available for thousands of hot subdwarfs thanks to the Kepler, K2, TESS and CHEOPS space missions (the harvest is continuing for these last two missions). By having experienced extreme mass loss on the RGB, these small stars (0.1-0.3 $R_{\odot}$) constitute excellent targets to address the question of the evolution of planetary systems directly after the first-ascent red giant branch. Hot subdwarfs also offer the potential to observationally constrain the existence of planetary remnants, i.e. planets that would have survived (even partially as a small, possibly disintegrating very close object) the engulfment in the envelope of their red giant host star. Not only does the small star size make possible the detection of small remnant objects, but the ejection of the envelope may actually be the cause of the survival of such remnants, by stopping the spiral-in inside the host star. Hot subdwarfs may therefore offer the outstanding opportunity to study the interior of giant planets, for which the exact structure is uncertain even for Jupiter \citep[][and refererences therein]{2017GeoRL..44.4649W}.

We first listed the hot subdwarfs observed by Kepler, K2, TESS and CHEOPS. We then performed injection-and-recovery tests for a selection of representative targets from Kepler, K2 and TESS, with the aim to determine which transiting bodies, in terms of object radius and orbital period, we will be able to detect in these light curves with our tools. For CHEOPS targets, given the filler nature of the observations (they are carried out when CHEOPS has no time-constrained or higher priority observations), we directly estimated the minimum planet size detectable from the SNR of the light curves, and computed afterwards which orbital periods are covered for a given phase coverage. For comparison purposes, we considered the same host star in all cases. 

Objects below $\sim$1$R_{\Earth}$ size are detectable (if existing and transiting, of course) for the shortest orbital periods (about 1~d and below) in most of Kepler, K2, and CHEOPS targets. Moon-sized values ($\sim$0.3 $R_{\Earth}$) are achievable in the best cases. Such performance of reaching sub-Earth-sized objects is obtained only for the very few brightest TESS data, as well as for stars with G $\lesssim 13$ observed for a significant number ($\gtrsim$ 6) of sectors. Altogether, we estimated that for about 250 targets we will be able to detect planets smaller than the Earth for orbital periods below 1~d, should they exist. Given the relatively high probability of transits for very close objects ($\approx$ 5\% at 1~d orbital period), our results demonstrate that we will actually be able to observationally determine if planets are able to survive the engulfment in the envelope of their host star. Hot subdwarfs represent a short phase of stellar evolution ($\sim$ 150 Myr for the core-He burning, i.e. EHB, phase, and about 10\% of that time for post-EHB evolution; \citealt{2016PASP..128h2001H}), which renders unlikely the formation of second-generation planets, in particular in light of the harsh environment for planet formation around a hot subdwarf. Migration of farther bodies that were not engulfed in the envelope of the red giant host would be possible for the oldest hot subdwarfs \citep{2018MNRAS.476.3939M}, although such a lifetime is likely too short for a complete circularization of the orbit. Dedicated computations will be needed, as carried out for the planets and remnants discovered around white dwarfs \citep[][and references therein]{2020MNRAS.492.6059V}.

Our tests also provided a series of representative results for the detection of farther and bigger planets. TESS targets will provide the most important cohort to the final goal of this project, which is to provide statistically significant occurrence rates of planets, as a function of object radius and orbital period, around hot subdwarfs. 

Our main pipeline for the search for transit events around hot subdwarfs, {\fontfamily{pcr}\selectfont SHERLOCK}, has already been successfully applied in a number of cases \citep{pozuelos2020a,demory2020}. However, there are several implementations under development which are especially relevant given the nature of our targets. The first improvement involves more efficient detrending for pulsating stars \cite[see e.g.][]{sowicka2017}, in particular high-frequency p-mode hot subdwarf pulsators, which have relatively high amplitudes that can hinder the detection of shallow transits. Secondly, we are including in {\fontfamily{pcr}\selectfont SHERLOCK} a model for comet-like tails of disintegrating exoplanets, which highly differ from the typical shape of transiting exoplanets \cite[see e.g.][]{brogi2012,rappaport2012,sanchis2015,2019MNRAS.482.5587K}.  

In case of identified transit event in the light curves successfully passing all the thresholds and the vetting process, we will need to confirm the signal and associate it to a planetary nature by scheduling follow-up observations. In order to confirm transit events in light curves, we will trigger for the deepest signals ($\gtrsim 2500$ ppm) observations with our Li\` ege TRAPPIST network \citep{2011Msngr.145....2J,2011EPJWC..1106002G}, which consists in two 0.6-m telescopes, at the La Silla (Chile) and Oukaïmeden (Morocco) observatories. For shallower transits we will directly use CHEOPS, provided the target has a sufficient visibility from the CHEOPS orbit. Once transits are confirmed, it will be necessary to discard a stellar, white dwarf or brown dwarf origin from RV measurements. We will first search for RV data in archives, open to the community (such as the ESO archives), or within the hot subdwarf community. We will write proposals on appropriate spectrographs when necessary. 

Finally, we will compute the occurrence rates of planets around hot subdwarfs by following a method similar to \citet{2018MNRAS.474.4603V,2019MNRAS.487..133W}. By comparing our results to these statistics for white dwarfs, to those for $\sim$0.8-2.3 $M_{\odot}$ main sequence stars that are the main progenitors of hot subdwarfs (e.g. \citealt{2011arXiv1109.2497M,2012ApJS..201...15H,2013ApJ...766...81F}), as well as for subgiants and RGB stars \citep{2008PASJ...60.1317S,2009A&A...505.1311D,2020arXiv200601277J}, we will be able to appreciate the effect of the RGB phase alone on the evolution of exoplanetary systems. 

\begin{acknowledgements}
We thank the anonymous referee for comments that improved the manuscript. The authors thank the Belgian Federal Science Policy Office (BELSPO) for the provision of financial support in the framework of the PRODEX Programme of the European Space Agency (ESA) under contract number PEA 4000131343. This work has been supported by the University of Liège through an ARC grant for Concerted Research Actions financed by the Wallonia-Brussels Federation. The authors acknowledge support from the Swiss NCCR PlanetS and the Swiss National Science Foundation. V.V.G. is a F.R.S.-FNRS Research Associate. M.G. is an F.R.S.-FNRS Senior Research Associate. St.C. acknowledges financial support from the Centre National d’Études Spatiales (CNES, France) and from the Agence Nationale de la Recherche (ANR, France) under grant ANR-17-CE31-0018. K.G.I. is the ESA CHEOPS Project Scientist and is responsible for the ESA CHEOPS Guest Observers Programme. She does not participate in, or contribute to, the definition of the Guaranteed Time Programme of the CHEOPS mission through which observations described in this paper have been taken, nor to any aspect of target selection for the programme. D.E. has received funding from the European Research Council (ERC) under the European Union’s Horizon 2020 research and innovation programme (project {\sc Four Aces}; grant agreement No 724427). This project has been carried out in the frame of the National Centre for Competence in Research PlanetS supported by the Swiss National Science Foundation (SNSF). G.B. acknowledges support from CHEOPS ASI-INAF agreement n. 2019-29-HH.0. A.J.M. acknowledges funding from the Swedish Research Council (starting grant 2017-04945) and the Swedish National Space Agency (career grant 120/19C). A.C.C. and T.G.W. acknowledge support from STFC consolidated grant number ST/M001296/1. A.B. was supported by the SNSA. M.F. gratefully acknowledge the support of the Swedish National Space Agency (DNR 65/19, 174/18). S.H. acknowledges CNES funding through the grant 837319. S.C.C.B. acknowledges support from FCT through FCT contracts nr. IF/01312/2014/CP1215/CT0004. S.G.S. acknowledge support from FCT through FCT contract nr. CEECIND/00826/2018 and POPH/FSE (EC). This work was supported by FCT - Fundação para a Ciência e a Tecnologia through national funds and by FEDER through COMPETE2020 - Programa Operacional Competitividade e Internacionalização by these grants: UID/FIS/04434/2019; UIDB/04434/2020; UIDP/04434/2020; PTDC/FIS-AST/32113/2017 \& POCI-01-0145-FEDER- 032113; PTDC/FIS-AST/28953/2017 $\&$ POCI-01-0145-FEDER-028953; PTDC/FIS-AST/28987/2017 $\&$ POCI-01-0145-FEDER-028987. O.D.S.D. is supported in the form of work contract (DL 57/2016/CP1364/CT0004) funded by national funds through FCT. B.-O.D. acknowledges support from the Swiss National Science Foundation (PP00P2-190080). B.N.B. acknowledges funding through the TESS Guest Investigator Program Grant 80NSSC21K0364. We acknowledge support from the Spanish Ministry of Science and Innovation and the European Regional Development Fund through grants ESP2016-80435-C2-1-R, ESP2016-80435-C2-2-R, PGC2018-098153-B-C33, PGC2018-098153-B-C31, ESP2017-87676-C5-1-R, MDM-2017-0737 Unidad de Excelencia “María de Maeztu”- Centro de Astrobiología (INTA-CSIC), as well as the support of the Generalitat de Catalunya/CERCA programme. The MOC activities have been supported by the ESA contract No. 4000124370. I.R. acknowledges support from the Spanish Ministry of Science and Innovation and the European Regional Development Fund through grant PGC2018-098153-B- C33, as well as the support of the Generalitat de Catalunya/CERCA programme. X.B., Se.C., D.G., M.F. and J.L. acknowledge their role as ESA-appointed CHEOPS science team members. D.G. gratefully acknowledges financial support from the CRT foundation under Grant No. 2018.2323 ``Gaseous or rocky? Unveiling the nature of small worlds''. P.F.L.M. acknowledges support from STFC research grant number ST/M001040/1. This  project  has  been  supported  by  the  Hungarian National Research, Development and Innovation Office (NKFIH) grants GINOP-2.3.2-15-2016-00003, K-119517,  K-125015, and the City of Szombathely under Agreement No.\ 67.177-21/2016. This paper includes data collected by the TESS mission. Funding for the TESS mission is provided by the NASA Explorer Program. Funding for the TESS Asteroseismic Science Operations Centre is provided by the Danish National Research Foundation (Grant agreement no.: DNRF106), ESA PRODEX (PEA 4000119301) and Stellar Astrophysics Centre (SAC) at Aarhus University. We thank the TESS team and staff and TASC/TASOC for their support of the present work. This work has made use of data from the ESA mission Gaia (\url{https://www.cosmos.esa.int/gaia}), processed by the Gaia Data Processing and Analysis Consortium (DPAC, \url{https://www.cosmos.esa.int/web/gaia/dpac/consortium}). Funding for the DPAC has been provided by national institutions, in particular the institutions participating in the Gaia Multilateral Agreement.
\end{acknowledgements}

\bibliographystyle{aa}
\bibliography{ms.bbl}

\appendix
\section{List of hot subdwarfs observed in the original Kepler field}
Table \ref{sd_Kepler}.
\begin{table*}
\scriptsize
\caption{\label{sd_Kepler}List of hot subdwarfs observed in the original Kepler field$^{*}$.}
\begin{center}
\begin{tabular}{lclccc}
\hline\hline
KIC & Class & Other name & Kp& Quarters (SC) & Quarters (LC)  \tabularnewline
\hline
\textbf{sdB pulsators} &&&&&	\tabularnewline					
9472174 &	sdB+dM	&2M1938+4603 &	12.3&	Q0, Q5-Q17.2	& All: Q0-Q17.2	\tabularnewline
2437937	& sdB&	B5 (NGC6791)&	13.9&	Q11.X	&Q11	\tabularnewline
3527751&	sdB &	J19036+3836& 	14.8	&Q2, Q5-Q17.2	& Idem SC	\tabularnewline
11558725&	sdB+WD	&J19265+4930 &	14.9&	Q3.3, Q6-Q17.2	& Q3, Q5-Q17.2	\tabularnewline
5807616	&sdB&	KPD 1943+4058	&15.0	&Q2.3, Q5-Q17.2&	Idem SC	\tabularnewline
10553698	&sdB&	J19531+4743 	&15.1&	Q4.1, Q8-Q10, Q12-Q14, Q16-Q17.2&	Q4-Q6, Q8-Q10, Q12-Q17.2	\tabularnewline
2697388	&sdB&	J19091+3756 	&15.4	&Q2.3, Q5-Q17.2	& Idem SC	\tabularnewline
7668647	&sdB+WD&	FBS1903+432 	&15.4&	Q3.1, Q6-Q17.2	&Q3.1, Q5-Q17.2	\tabularnewline
10001893	&sdB&	J19095+4659& 	15.8&	Q3.2, Q6-Q17.2	&Q3.2, Q5-Q17.2	\tabularnewline
10139564	&sdB&	J19249+4707 &	16.1&	Q2.1, Q5-Q17.2&	Idem SC	\tabularnewline
8302197	&sdB&	J19310+4413 &	16.4	&Q3.1, Q5-Q17.2 except Q12	&idem SC	\tabularnewline
7664467	&sdB&	J18561+4319 &	16.4	&Q2.3, Q5-Q17.2 except Q12	&idem SC	\tabularnewline
10670103	&sdB&	J19346+4758 &	16.5	&Q2.3, Q5-Q17.2&	Idem SC\tabularnewline
11179657	&sdB+dM&	J19023+4850& 	17.1&	Q2.3, Q5-Q17.2 except Q8 and Q12&	idem SC	\tabularnewline
2991403	&sdB+dM	&J19272+3808	&17.1&	Q1, Q5-Q17.2&	Idem SC	 \tabularnewline
2991276	&sdB&	J19271+3810 	&17.4&	Q2.1, Q6-Q17.2 except Q12&	Idem SC	\tabularnewline
2569576	&sdB&	B3 (NGC6791)&	18.1&	Q11.3, Q14-Q17.2	&Q11, Q14-Q17.2	\tabularnewline
2438324	&sdB+dM&	B4 (NGC6791)	&18.3	&Q6-Q17.2	&Idem SC	\tabularnewline

\textbf{sdB/sdOB non pulsators}&&&&&	\tabularnewline	
6848529	&sdB+?&	BD +42 3250 &	10.7&	Q0	&All: Q0-Q17.2 	\tabularnewline	
1868650	&sdB+dM	&KBS 13	&13.4&	Q1&	All: Q0-Q17.2 	\tabularnewline	
9543660	&sdOB	&&	13.8	&Q1	&Q1-Q17, except Q7 and Q11	\tabularnewline	
10982905	&sdB+F/G&	J19405+4827& 	14.1	&Q2.1&	Q2-Q10	\tabularnewline	
6188286	&sdOB	&&	14.2&	Q2.3	&Q2, Q6-Q8, Q14-Q16	\tabularnewline	
8054179	&He-sdOB	&&	14.4&	Q3.1, Q6&	Q3.1, Q4-Q17.2 except Q11 and Q12	\tabularnewline	
7975824	&sdOB+WD&	KPD 1946+4340	&14.6&	Q1, Q5-Q12&	Q1, Q5-Q17.2	\tabularnewline	
10449976	&He-sdOB&&		14.9	&Q3.2	& Q3, Q5-Q9	\tabularnewline	
3353239	&sdB	&&	15.2&	Q4.1	&Q4-Q5, Q7-Q9, Q13-Q17	\tabularnewline	
10593239	&sdB+F/G	&J19162+4749 &	15.3&	Q2.3	& Q2, Q5-Q17.2	\tabularnewline	
2569583	&sdB&	B6 (NGC6791)	&15.4	&Q11.2	&Q11	\tabularnewline	
7104168	&sdB	&&	15.5	&Q3.1	&Q3, Q5-Q9	\tabularnewline	
10149211	&sdB+?	&&	15.5	&Q4.2	&Q4-Q17.2	\tabularnewline	
10789011	&sdOB	&&	15.5&	Q3.2	&Q3, Q5-Q10	\tabularnewline	
11350152	&sdB+F/G	&&	15.5	&Q3.1	&Q3, Q5-Q10	\tabularnewline	
7434250	&sdB+?	&J19135+4302 &	15.5	&Q2.3	&Q2, Q5-Q17.2	\tabularnewline	
2020175	&sdB	&&	15.5	&Q3.1	&Q3, Q5-Q10, Q13-Q17.2	\tabularnewline	
12021724	&sdB+WD?	&&	15.6&	Q4.2&	Q4-Q10	\tabularnewline	
3343613	&He-sdOB		&&15.7	&Q3.2&	Q3, Q5-Q10	\tabularnewline	
5938349	&sdB	&&	16.1	&Q3.2	&Q3, Q10	\tabularnewline	
6614501&	sdB+WD?&&		16.1	&Q3.3, Q5, Q6, Q8-Q10	&Q3.3, Q5-Q17.2	\tabularnewline	
9211123	&sdB	&&	16.1	&Q3.3	&Q3, Q5-Q10, Q13-Q17.2	\tabularnewline	
9957741	&He-sdOB&&		16.1	&Q2.1	&Q2, Q6-Q9	\tabularnewline	
2304943	&sdB	&&	16.2	&Q3.3	&Q3, Q10	\tabularnewline	
8496196	&sdOB	&&	16.4	&Q2.3	&Q2, Q6-Q10	\tabularnewline	
8874184	&sdB+?	&&	16.5	&Q4.1	&Q4-Q10, Q13-Q17.2	\tabularnewline	
8022110	&sdB	&&	16.5	&Q2.3	&Q2, Q6-Q10, Q13-Q17.2	\tabularnewline	
6878288	&He-sdOB+?	&&	16.7	&Q3.1	&Q3, Q5-Q10	\tabularnewline	
6522967	&sdB	&&	16.9&	Q3.2	&Q3, Q10	\tabularnewline	
7799884	&sdB	&&	16.9	&Q4.1	&Q4.1	\tabularnewline	
10462707	&sdB+WD?	&&	16.9	&Q4.1&	Q4.1, Q10	\tabularnewline	
11400959	&sdB	&&	16.9&	Q4.1&	Q4.1	\tabularnewline	
10784623	&sdB	&&	17.0	&Q10&	Q4-Q10 except Q8	\tabularnewline	
10961070	&sdOB	&&	17.0&	Q4.2	&Q4.2	\tabularnewline	
3527028	&sdB+?	&&	17.1	&Q4.2	&Q4-Q10	\tabularnewline	
5340370	&sdB+?&&		17.1&	Q4.2&	Q4, Q10	\tabularnewline	
9569458	&sdB	&&	17.2&	Q1&	Q1	\tabularnewline	
8889318&	sdB	&&	17.2&	Q2.3	&Q2.3, Q13-Q17.2	\tabularnewline	
9408967	&He-sdOB	&&	17.2&	Q2.3&	Q2.3, Q10	\tabularnewline	
4244427	&sdB	&&	17.3&	Q2.1, Q6-Q10	&Q2.1, Q6-Q17.2 except Q12	\tabularnewline	
8142623	&sdB+?	&J18427+4404 &	17.3&	Q1	&Q1, Q5-Q17.2	\tabularnewline	
11357853	&sdOB	&&	17.4&	Q2.1&	Q2.1	\tabularnewline	
3527617&	He-sdOB&	&	17.5&	Q2.2&	Q2.2	\tabularnewline	
3729024&	sdB&&		17.6&	Q2.2&	Q2.2	\tabularnewline	
9095594	&sdB&	&	17.7&	Q3.2&	Q3.2	\tabularnewline	
5342213	&sdOB&	&	17.7	&Q2.2&	Q2.2, Q14-Q16	\tabularnewline	
10661778	&sdB&&		17.7	&Q2.3, Q6-Q10	&Q2.3, Q6-Q17.2 except Q11 and Q12	\tabularnewline	
\textbf{sdO non pulsators}&&&&&	\tabularnewline	
7755741&	sdO	&&	13.7&	Q1&	Q1-Q17	\tabularnewline	
9822180	&sdO+F/G	&&	14.6&	Q2.1, Q6&	Q2.1, Q6-Q10	\tabularnewline	
7353409	&sdO&&		14.7&	Q2.2, Q5&	Q2.2, Q5-Q9	\tabularnewline	
10207025	&He-sdO	&&	15.0&	Q3.3	&Q3.3, Q5-Q9	\tabularnewline	
7335517	&sdO+dM	&&	15.7&	Q3.2, Q6&	Q3.2, Q5-Q17.2	\tabularnewline	
2297488	&sdO+F/G		&&17.2&	Q1&	Q1	\tabularnewline	
2303576	&He-sdO+?	&&	17.4&	Q3.3, Q6&	Q3.3, Q6-Q17.2	\tabularnewline	

\hline\hline
\end{tabular}
\end{center}
$^{*}$Commissioning (9.7 days starting 2 May 2009): Q0; Survey phase: Q1: 33.5 d (12 May-14 June 2009), Q2, Q3, and Q4: about 90 days each, divided in 3, i.e. monthly surveys; Rest of the mission: Q5 to Q16: about 90 days each; mission stopped at Q17.2 (11 May 2013).								
\end{table*}

\section{List of hot subdwarfs observed in the K2 fields}
Table \ref{sd_K2}.

\clearpage
\onecolumn
{\footnotesize
\begin{longtable}{lllccc}

\caption{\label{sd_K2} List of hot subdwarfs observed in the K2 fields$^{\dagger}$.}
\\\hline\hline
KIC & Class & Other name & Kp& Campaign (SC) & Campaign (LC)  \tabularnewline
\hline
\endfirsthead
\caption{\label{sd_K2} (Continued).}
\\\hline\hline
KIC & Class & Other name & Kp& Campaign (SC) & Campaign (LC)  \tabularnewline
\hline
\endhead
\hline
\endfoot

\multicolumn{6}{l}{\textbf{sdB pulsators}} \tabularnewline				
220641886&	sdB&	HD 4539&	10.40	&8&	8		\tabularnewline											
228755638&	sdB+dM	&HW Vir&	10.76&	10 (101-102)&	10 (101-102)		\tabularnewline					
211623711&	He-sdB&	UVO 0825+15&	11.89&	5;18	&5;18		\tabularnewline										
220376019&	sdB+WD&	PG 0101+039	&12.11	&8	&8	\tabularnewline														
220422705&	sdB+G&	PG 0039+049	&12.87	&8&	8	\tabularnewline														
249942493&	sdB	&EC 15103-1557&	12.89&	15	&15	\tabularnewline											
211779126&	sdB	&2M0856+1701	&12.92	&5;18&	5;18		\tabularnewline							
246387816&	sdB+dM&	EQ Psc&	12.92&	12	&12					\tabularnewline			
246023959&	sdB+dM	&PHL 457&	13.04&	12	&12							\tabularnewline
211881419&	iHe-sdB&	PG 0848+186	&13.30&	16;18&	5;16;18								\tabularnewline			
201203416&	sdB&	PG 1156-037	&13.46	&10 (101-102)	&10 (101-102)	\tabularnewline								
248411044&	sdB	&UY Sex	&13.56	&14	&14			\tabularnewline								
246141920&	sdB	&PHL 531&	13.99	&12	&12			\tabularnewline										
211433013&	sdB+WD&	LT Cnc	&14.02&	16&	16		\tabularnewline													
211765471&	sdB+WD	&HZ Cnc	&14.04&	5;16;18	&5;16;18	\tabularnewline														
220614972&	sdB+F&	PG 0048+091	&14.29	&8&	8		\tabularnewline											
211392098&	sdB+MS	&SDSS J082517.99+113106.3&	14.34&	18	&5;18		\tabularnewline													
211437457&	sdB	&PG 0902+124	&14.73&	16&	16	\tabularnewline														
246683636&	sdB+dM&	V1405 Ori	&15.07&	13	&13	\tabularnewline										
248368659&	sdB+WD	&VPHAS J181343.0-213843.9&	15.10&	9 (91-92)	&9 (91-92)		\tabularnewline													
212508753&	sdB+F7&	PG 1315-123&	15.13	&6;17	&6;17			\tabularnewline									
211823779&	sdB+F1	&SDSS J082003.35+173914.2&	15.22	&5;18	&5;18		\tabularnewline											
212475716&	sdB+MS	&EC 13356-1300&	15.24&	17	&17			\tabularnewline												
211696659&	sdB+WD	&SDSS J083603.98+155216.4	&15.50	&5;18&	5;18		\tabularnewline												
212707862&	sdB	&SDSS J135544.71-080354.3	&15.55&	6;17	&6;17		\tabularnewline												
212204284&	sdB	&PG 0843+246&	15.64	&16	&16					\tabularnewline										
246283223&	sdB	&HE 2307-0340	&15.66	&12	&12					\tabularnewline										
248368658&	sdB&		&15.70&	9 (91-92)	&9 (91-92)	\tabularnewline														
218717602&	sdB	&	&15.76&	7	&7		\tabularnewline												
211938328&	sdB+F6&	LB 378	&15.78&	5;18	&5;18			\tabularnewline											
218366972&	sdB+WD	&	&15.94&	7	&7							\tabularnewline						
201206621&	sdB+WD	&PG 1142-037&	15.99&	1	&1	\tabularnewline												
212487276&	sdB	&EC 13359-1245&	16.23&	17	&17	\tabularnewline														
217280630&	sdB	&	&16.33&	7	&7			\tabularnewline									
215776487&	sdB	&	&16.35&	7&	7	\tabularnewline												
203948264&	sdB	&	&16.70&	2	&2		\tabularnewline										
246373305&	iHe-sdB&	PHL 417&	16.88	&12	&12	\tabularnewline											
251668197&	sdB	&EC 15094-1725	&17.00	&15	&15	\tabularnewline											
229002689&	sdB	&SDSS J122057.48-012642.3	&18.65&	10 (101-102)	&10 (101-102)	\tabularnewline														
220188903& sdB+WD&	PB 6373&	14.91&	no data	&8 	\tabularnewline														
230195595&	sdB	&&	15.59&	no data&	11	\tabularnewline			
\tabularnewline
\tabularnewline												
\multicolumn{6}{l}{\textbf{sdB/sdOB non pulsators, single}}	\tabularnewline	
234319842&	sdB	&	&12.97&	11 (111-112)	&11 (111-112)		\tabularnewline															
60017832&	sdB	&PG 2349+002&	13.27&	T&		\tabularnewline																
211708181&	sdB	&GALEX J081233.6+160121	&13.77&	5&	5		\tabularnewline															
227389858&	sdB	&&	13.79&	11 (111-112)&	11 (111-112)		\tabularnewline															
246230928&	sdB	&PHL 529&	13.93&	12	&12			\tabularnewline														
206535752&	sdB&	PHL 358&	13.99&	3	&3		\tabularnewline															
201648341&	sdB&	PG 1214+031	&14.04&	10 (101-102)&	10 (101-102)	\tabularnewline																
217204898&	sdB	&&	14.26&	7&	7		\tabularnewline															
246643895&	sdB&	HS 0446+1344&	14.50&	13	&13		\tabularnewline															
212722777&	sdB&	PG 1330-074	&14.93&	17&	17		\tabularnewline															
211727748&	sdB&	PG 0838+165	&14.99&	5;16&	5;16		\tabularnewline															
206073023&	sdB&	BPS CS 29512-38	&15.00&	3&	3		\tabularnewline															
210837690&	sdB&&		15.11&	4&	4				\tabularnewline													
212498842&	sdB&	EC 13162-1229&	15.26&	6&	6		\tabularnewline															
212465180&	sdB&	EC 13265-1313&	15.56&	6&	6		\tabularnewline															
212160066&	sdB&	SDSS J082445.68+231520.3&	15.57&	18&	5;18	\tabularnewline																
246901153&	sdB&	KUV 04369+1640&	15.70&	13&	13		\tabularnewline															
249601610&	sdB&	EC 15050-2017&	15.71&	15&	15		\tabularnewline															
246980092&	sdB&	KUV 04482+1727&	15.74&	13&	13		\tabularnewline															
218148570&	sdB&&	15.74&	7&	7				\tabularnewline													
228914323&	sdB	&PG 1249-028&	15.76&	10 (101-102)&	10 (101-102)		\tabularnewline															
228682488&	sdB&	SDSS J085217.70+211637.4&	16.00&	16&	16			\tabularnewline														
212818294&	sdB&	PG 1356-047	&16.15&	6;17&	6;17	\tabularnewline																
248422838&	sdB&	PG 1032+007&	16.27&	14&	14		\tabularnewline															
214515136&	sdB	&&	16.30&	7&	7		\tabularnewline															
251603936&	sdB&	SDSS J131916.15-011405.0&	16.69&	17	&17	\tabularnewline																
201531672&	sdB	&SDSS J112757.48+010044.2&	16.89&	1&	1	\tabularnewline																
251457058&	sdB&	SDSS J105428.85+010514.7&	17.10&	14&	14		\tabularnewline															
246371369&	sdB	&PB 5212&	17.11&	12	&12				\tabularnewline													
211552072&	sdB&	SDSS J084556.85+135211.3&	17.50&	16	&16		\tabularnewline															
212567176&	sdB&	HE 1309-1102&	17.65&	6&	6		\tabularnewline															
249585191&	sdB&	EC 15064-2029&	17.95&	15&	15		\tabularnewline															
248840987&	sdB&	SDSS J102050.99+114024.3&	18.15&	14&	14		\tabularnewline
248810568&	sdOB&	SDSS J110055.94+105542.3&	14.22&	14&	14		\tabularnewline															
246997679&	sdOB&	KUV 05109+1739&	14.58&	13&	13			\tabularnewline														
211421561&	sdOB&	SDSS J090042.68+115749.9&	14.90&	16&	16		\tabularnewline															
220265912&	sdOB&	PG 0055+016	&15.19&	8&	8		\tabularnewline	
249700050&	sdOB&	EC 15059-1902&	15.65&	15&	15		\tabularnewline			
206240954&	sdOB&	SDSS J220337.88-090733.5&	16.31&	3&	3	\tabularnewline	
210731139&	sdOB&	SDSS J032427.24+184918.2&	16.37&	4&	4		\tabularnewline															
246087406&	sdOB&	PB 7470&	16.46&	12	&12		\tabularnewline															
206186190&	sdOB&	BPS CS 22886-65	&16.49&	3&	3		\tabularnewline															
251605347&	sdOB&	SDSS J133611.02-011156.0&	18.69&	17	&17		\tabularnewline															
246745570&	He-sdB	&KUV 04456+1502&	15.68&	13&	13		\tabularnewline															
211920209&	He-sdB&	PG 0850+192&	16.39&	18&	5; 16; 18	\tabularnewline																
249770424&	He-sdOB	&GALEX J152332.2-181726	&14.00	&15&	15		\tabularnewline															
211495446&	He-sdOB&	PG 0838+133	&14.03	&5;16&	5;16	\tabularnewline																
248748173&	He-sdOB&	PG 1033+097&	16.38&	14&	14			\tabularnewline														
248761152&	He-sdOB	&PG 1045+100&	17.09&	14&	14			\tabularnewline														
248915544&	He-sdOB	&SDSS J103806.64+134412.1&	17.21&	14	&14		\tabularnewline															211841249&	sdB	&SDSSJ082734.96+175356.0&	14.64&	5;18	&5;18		\tabularnewline	
250083298&	sdB	&EC15203-1418&	17.34&	15	&15		\tabularnewline	
\tabularnewline
\multicolumn{6}{l}{\textbf{sdB/sdOB non pulsators, in binaries}}	\tabularnewline
220468352&	sdB+F&	PB 6355&	13.01	&8&	8	\tabularnewline												
251377113&	sdB+F/G&	SDSS J090827.24+231417.9&	13.53&	16&	16	\tabularnewline														
211499370&	sdB+F/G/K&	SDSS J082556.80+130753.5&	14.60	&5&	5;18\tabularnewline															
218637228&	sdB+F/G	&&	14.79&	7&	7			\tabularnewline												
227441033&	sdB+F/G	&&	15.10&	11 (111-112)&	11 (111-112)\tabularnewline															
216924452&	sdB+F/G	&&	15.53	&7&	7	\tabularnewline														
250121838&	sdB+F/G/K&	EC 15365-1350&	15.74	&15	&15		\tabularnewline													
246151922&	sdB+G9&	HE 2322-0617&	15.74&	12;19	&12;19\tabularnewline															
212630158&	sdB+F/G	&&	15.75&	6&	6	\tabularnewline														
246868556&	sdB+F/G	&GALEX J050252.2+162647&	15.78&	13&	13	\tabularnewline														
246864591&	sdB+F/G/K&	KUV 04571+1620&	15.98&	13	&13	\tabularnewline														
211910684&	sdB+F/G	&PG 0906+191&	15.99&	16&	16		\tabularnewline													
212108396&	sdB+F/G	&SDSS J082447.30+221112.9&	16.02&	5	&5;18\tabularnewline															
211400847&	sdB+F/G&	SDSS J084447.00+113910.0&	16.43&	5&	5;18\tabularnewline															
212003762&	sdB+F/G	&SDSS J081406.79+201901.7	&16.51&	18&	18	\tabularnewline														
212137838&	sdB+F/G	&Ton 920&	16.54&	5&	5				\tabularnewline											
250152590&	sdB+F/G/K&	LB 889&	17.13&	15&	15			\tabularnewline												
248467942&	sdB+F/G&	SDSS J103022.07+020524.3&	17.24&	14&	14	\tabularnewline														
211732575&	sdB+F/G&	SDSS J082426.51+162145.1&	17.68&	18&	18	\tabularnewline														
251583165&	sdB+F/G&	SDSS J131932.19-014131.2&	18.24&	17&	17	\tabularnewline														
212866280&	sdB+F/G&	SDSS J133701.51-031732.2&	18.27&	17&	17	\tabularnewline														
212410755&	sdB+WD&	EC 13332-1424&	13.46&	6&	6		\tabularnewline													
201535046&	sdB+?&	PG 1049+013&	14.44&	14&	14	\tabularnewline														
251372905&	sdOB+F/G&	SDSS J091216.06+225452.7&	15.30&	16&	16\tabularnewline															
211904152&	sdOB+F/G&	PG 0912+189&	15.93&	16&	16			\tabularnewline												
248767552&	sdOB+WD?&	SDSS J101833.11+095336.1&	14.97&	14&	14	\tabularnewline														
246877984&	sdOB+WD&	KUV 05053+1628&	16.11&	13&	13			\tabularnewline	
\tabularnewline
\multicolumn{6}{l}{\textbf{sdO non pulsators}}	\tabularnewline	
212762631&	sdO&	PG 1355-064&	13.76&	6&	6;17\tabularnewline																
220179214&	sdO&	GD 934&	14.93	&8	&8		\tabularnewline														
248520995&	sdO&	SDSS J110053.55+034622.8&	17.25&	14&	14	\tabularnewline															
211517387&	sdO&	SDSS J082944.74+132302.5&	17.32&	5&	5	\tabularnewline															
249862817&	sdO&	EC 15447-1656&	18.05&	15&	15	\tabularnewline															
228821386&	He-sdO&	PG 1220-056&	14.86&	10 (101-102)&	10 (101-102)\tabularnewline																
249867379&	He-sdO&	EC 15348-1652&	15.35&	15	&15	\tabularnewline															
205247324&	He-sdO&	&	16.01&	2&	2	\tabularnewline															
201640895&	He-sdO&	SDSS J110215.45+024034.2&	17.60&	14	&14	\tabularnewline															
228960704&	He-sdO&	SDSS J123821.48-021211.4&	18.49&	10 (101-102)	&10 (101-102)\tabularnewline										\tabularnewline						
\multicolumn{6}{l}{\textbf{Misc., in LC only}}	\tabularnewline	
															
201150341&	sdB&	HE 1140-0500&	14.50&	no data	&1	\tabularnewline														
214958569&	sdB	&&	15.70&	no data	&7	\tabularnewline														
216775790&	sdB	&&	16.50&	no data&	7	\tabularnewline														
201236182&	sdB	&PG 1154-031&	16.59&	no data&	1	\tabularnewline														
211720816&	sdB	&SDSS J083901.50+161148.0&	16.71&	no data	&5; 16; 18\tabularnewline															
211594465&	sdB	&SDSS J081931.22+142756.1&	17.19&	no data	&5; 18\tabularnewline															
248912731&	sdB	&SDSS J103832.41+133848.3&	17.44&	no data	&14\tabularnewline															
201201339&	sdB	&SDSS J112757.48+010044.2&	17.50&	no data&	1\tabularnewline															
201590024&	sdB&	SDSS J113418.00+015322.1&	17.65&	no data	&1\tabularnewline															
201698091&	sdB&	SDSS J114821.29+033625.7&	17.70&	no data&	1\tabularnewline															
229021782&	sdB&	SDSS J125410.86-010408.3&	17.72&	no data &	10	\tabularnewline														
228682339&	sdB&	SDSS J082824.20+212556.7&	17.73&	no data	&5; 16; 18	\tabularnewline														
251457060&	sdB&	SDSS J104725.10+010847.2&	17.80&	no data	&14	\tabularnewline														
248783069&	sdB&	SDSS J104620.14+101629.7&	18.65&	no data&	14\tabularnewline															
251410019&	sdB&	SDSS J085809.09+252134.6&	18.87&	no data &	16	\tabularnewline	
201424163&	sdB+WD&	PG 1136-003&	15.96&	no data	&1	\tabularnewline															
228682347&	sdB+WD&	SDSS J083139.68+162316.4&	17.91&	no data&	5	\tabularnewline															
248783744&	sdB+WD&	SDSS J103218.40+101725.8&	18.82&	no data	&14	\tabularnewline															
211460944&	sdB+WD ?&	SDSS J084556.85+135211.3&	15.36&	no data&	16	\tabularnewline	
228796212&	sdB&	SDSS J124446.64-065625.8&	18.83&	no data&	10	\tabularnewline														
211991114&	sd+F/G&	Ton 914&	15.10&	no data&	5; 18		\tabularnewline													
211930840&	He-sdB&	SDSS J091512.06+191114.6&	19.13&	no data	&16	\tabularnewline														
201734164&	sdOA&	PG 1110+045&	14.84&	no data&	1		\tabularnewline													
213545287&	sdOB&	GALEX J191509.0-290311&	15.00&	no data	&7	\tabularnewline														
201924421&	sdOB&	SDSS J113218.41+075103.0&	17.20&	no data	&1\tabularnewline															
228682323&	sdOB&	SDSS J082110.89+183924.1&	17.84&	no data	&5	\tabularnewline														
212034957&	sdOB&	SDSS J090302.39+205008.9&	18.62&	no data	&16	\tabularnewline														
215669184&	He-sdOB&	GALEX J193323.6-234553&	15.00&	no data	&7	\tabularnewline														
201802867&	He-sdOB&	SDSS J111633.29+052507.9&	17.80	&no data&	1	\tabularnewline														
251383153&	He-sdOB&	SDSS J091044.90+234044.6&	18.27&	no data&	16	\tabularnewline														
229155531&	He-sdOB&	SDSS J121643.72+020835.9&	18.73&	no data&	10\tabularnewline															
251357585&	He-sdOB&	SDSS J092245.79+214238.9&	19.01&	no data&	16	\tabularnewline													
211602914&	sd&	SDSS J082959.28+143441.8&	15.64&	no data& 	5; 18\tabularnewline
213716821 & sdO &GALEX J192041.4-282939&13.40&no data&7	\tabularnewline
246735349 & He-sdO &KUV 04402+1455&13.97&no data&	13\tabularnewline
214453765 & sdO &GALEX J191158.1-262712&15.30&no data& 7	\tabularnewline
216452306 &He-sdO &&16.40&no data& 7	\tabularnewline

231422890 &sdO &&17.07&no data&	11\tabularnewline
201418759 & sdO &SDSS J111438.57-004024.3&18.10&no data&1	\tabularnewline
201843731 &sdO &SDSS J115009.48+061042.1&18.10&no data&1	\tabularnewline
211559083 &sdO &SDSS J084421.10+135807.6&18.18&no data&16	\tabularnewline
228682365 & He-sdO &SDSS J083747.23+194955.9&18.60&no data&5	\tabularnewline
216747137 & sdO+dM &2MASS J18521800-2147506&13.87&no data&7	\tabularnewline
217750936 & sdO+dM&&16.70&no data&7	\tabularnewline
\hline\hline
										
\multicolumn{6}{l}{$^{\dagger}$T: Engineering test from 4 to 13 Feb 2014; Campaign 0 (8 March-27 May 2014) to 18 (12 May - 2 July 2018),} \tabularnewline 
\multicolumn{6}{l}{\url{https://keplerscience.arc.nasa.gov/k2-fields.html}}
\end{longtable}
}
\twocolumn
\section{List of hot subdwarfs observed in TESS primary mission}
\label{CDS_primary}
i.e., Sector 1 to 26. {\url{https://github.com/franpoz/Hot-Subdwarfs-Catalogues}}
\section{List of hot subdwarfs observed in TESS extended mission}
\label{CDS_extended}
Sector 27 to Sector 32. {\url{https://github.com/franpoz/Hot-Subdwarfs-Catalogues}}
\end{document}